\documentclass[12pt,preprint,numberedappendix]{aastex}
\usepackage{epsf}

\newcommand{\tx}[1]{\textrm{#1}}

\newcommand{\ltsima}{$\buildrel<\over\sim$}
\newcommand{\lapprox}{\lower.5ex\hbox{\ltsima}}
\newcommand{\kms}{km~$\tx{s}^{-1}$}
\newcommand{\dv}{$r^{1/4}\,$}

\newcommand{\resecb}{$r_{\tx{e4}}$}

\newcommand{\resecv}{$r_{\tx{e6}}$}

\newcommand{\reseci}{$r_{\tx{e8}}$}

\newcommand{\resecz}{$r_{\tx{e9}}$}

\newcommand{\be}{\begin{equation}}
\newcommand{\ee}{\end{equation}}

\newenvironment{inlinefigure}{
\def\@captype{figure}
\noindent\begin{minipage}{0.999\linewidth}\begin{center}}
{\end{center}\end{minipage}\smallskip}

\slugcomment{ApJ, submitted}

\shorttitle{Spectroscopy of Distant Spheroids in GOODS-N}
\shortauthors{Treu et al.}

\begin{document}

\title{The Assembly History of Field Spheroidals: Evolution of Mass-to-light Ratios 
and Signatures of Recent Star Formation}

\author{Tommaso Treu\altaffilmark{1,2,3}, Richard
S. Ellis\altaffilmark{3}, Ting Xi Liao\altaffilmark{3}, Pieter G. van
Dokkum \altaffilmark{4,3}, Paolo Tozzi\altaffilmark{5}, Alison Coil\altaffilmark{5},
Jeffrey Newman \altaffilmark{6,2}, Michael C. Cooper\altaffilmark{6}, Marc
Davis\altaffilmark{6}}

\altaffiltext{1}{Department of Physics and Astronomy,
University of California at Los Angeles, Los Angeles, CA 90095; ttreu@astro.ucla.edu}
\altaffiltext{2}{Hubble Fellow}
\altaffiltext{3}{Caltech, Astronomy, 105-24, Pasadena, CA 91125,
USA}
\altaffiltext{4}{Department of Astronomy, Yale University, New
Haven, CT 06520} 
\altaffiltext{5}{INAF-Osservatorio Astronomico di Trieste
via G.B. Tiepolo 11, 34131 Trieste - Italy}
\altaffiltext{6}{Department of Astronomy, UC Berkeley, CA 94720}

\footnotetext{\footnotesize Using data obtained with the Hubble
Space Telescope operated by AURA for NASA and the W.M. Keck
Observatory on Mauna Kea, Hawaii. The W.M. Keck Observatory is
operated as a scientific partnership among the California
Institute of Technology, the University of California and NASA and
was made possible by the generous financial support of the W.M.
Keck Foundation.}


\begin{abstract}

We present a comprehensive catalog of high signal-to-noise spectra
obtained with the DEIMOS spectrograph on the Keck II telescope for a
sample of F850LP$<$22.43 (AB) field spheroidal (E+S0s; 163) and bulge
dominated disk (61) galaxies in the redshift range $0.2<z<1.2$,
selected on the basis of visual morphology from the northern field of
the Great Observatories Origins Deep Survey (GOODS-N). We discuss
sample selection, photometric properties and spectral reduction.  We
derive scale-lengths, surface brightnesses and photometric
inhomogeneities from the ACS data, and redshifts, stellar velocity
dispersions, [\ion{O}{2}] and H$\delta$ equivalent widths from the
Keck spectroscopy. Using the published 2Ms Chandra X-ray catalog we
identify active galactic nuclei to clarify the origin of emission
lines seen in the Keck spectra. Only 2/13 [\ion{O}{2}] emitting
early-type galaxies are identified as secure AGN on the basis of their
X-ray emission.  Contrary to earlier suggestions, we find that most
spheroidals containing `blue cores' are not associated with
non-thermal nuclear activity.  We examine the zero point, tilt and
scatter of the Fundamental Plane (FP) as a function of redshift and
morphological properties, carefully accounting for
luminosity-dependent biases via Montecarlo simulations. The evolution
of the overall FP can be represented by a mean change in effective
mass-to-light ratio given by $<d \log (M/L_{\rm
B})/dz>=-0.72^{+0.07}_{-0.05}\pm0.04$.  However, this evolution
depends significantly on the dynamical mass, being slower for larger
masses as reported in a previous letter. In addition, we separately
show the intrinsic scatter of the FP increases with redshift as
d(rms($M/L_{\rm B}$))/dz=$0.040\pm0.015$.  Although these trends are
consistent with single burst populations which formed at $z_f>2$ for
high mass spheroidals and $z_{f}\sim 1.2$ for lower mass systems, a
more realistic picture is that most of the stellar mass formed in all
systems at $z>2$ with subsequent activity continuing to lower
redshifts ($z<1.2$). The fraction of stellar mass formed at recent
times depend strongly on galactic mass, ranging from $<1$\% for masses
above 10$^{11.5} M_{\odot}$ to 20-40\% below 10$^{11}$
M$_{\odot}$. Independent support for recent activity is provided by
spectroscopic ([\ion{O}{2}] emission, H$\delta$) and photometric (blue
cores and broad-band colors) diagnostics.  Via the analysis of a large
sample with many independent diagnostics, we are able to reconcile
previously disparate interpretations of the assembly history of field
spheroidals. We discuss the implications of this measurement for the
determination of the evolution of the number density of E+S0s
galaxies, suggesting number density evolution of the morphologically
selected population has occurred since $z\sim$1.2.
\end{abstract}


\section{Introduction}

The assembly history of field spheroidal galaxies (defined here to
include both ellipticals and lenticulars, in short E+S0s) remains an
important issue in galaxy formation and particularly in the interface
between theory and observation (see Treu 2004, McCarthy 2004 for
recent reviews). The overall assembly rate of stellar mass has been
constrained both by measures of the cosmic star formation history
(Madau et al.\ 1996, Blain et al.\ 1999, Steidel et al.\ 1999, Bunker
et al.\ 2004) and, directly, via photometric estimates of the stellar
mass density as a function of redshift (Brinchmann \& Ellis 2000,
Cohen 2001, Papovich et al.\ 2001, Rudnick et al.\ 2003; Fontana et
al.\ 2004). However, such global constraints integrate over important
details relating to the formation history of individual morphological
types.

A key prediction of current semi-analytical models based on the cold
dark matter paradigm (e.g. Kauffmann 1996, Baugh et al.\ 1998,
Somerville \& Primack 1999) is the late assembly of massive spheroids
from smaller fragments with significant consequences for the abundance
and mass distribution of systems at $z\simeq$1-2.  Although the actual
rate of assembly is sensitive to details of the algorithms used in the
semi-analytical codes (see comparisons in Benson et al.\ 2003), the
observational data has provided only mixed support for the evolution
expected.

The simplest evolutionary test is based on measuring the density of
spheroids at various redshifts. Brinchmann \& Ellis (2000), and more
recently Bundy, Ellis \& Conselice (2005), found only a modest decline
in the {\it stellar mass density} to $z\simeq$1 using
morphologically-defined samples (see also Im et al.\ 2002, Cross et
al.\ 2004). These results appear to contrast with the claim by Bell et
al.\ (2003), based on color-selected samples, that the {\it luminosity
density} associated with red galaxies is fairly constant to $z\simeq$1
suggesting a marked drop in the volume density of such objects. The
utility of these tests depends sensitively on corrections for
incompleteness and cosmic variance.

An independent approach involves estimating the ongoing mass accretion
rate for spheroidals, either via the presence of photometric
abnormalities such as `blue cores' (Menanteau et al.\ 2001a,b, 2004,
2005), spectroscopic signatures of recent activity (Trager et al.\
2000; Treu et al.\ 2002; van Dokkum \& Ellis 2003) or deviations from
passive trends predicted in the evolution of the Fundamental Plane
(Djorgovski \& Davis 1987, Dressler et al. 1987; hereafter FP; Treu et
al.\ 1999, 2001b, 2002; van Dokkum \& Ellis 2003; Gebhardt et al.\
2003; van de Wel et al.\ 2004; Treu \& Koopmans 2004; see also
Kochanek et al.\ 2000; Rusin et al.\ 2003a; van de Ven et al.\ 2003;
Rusin \& Kochanek 2005). The degree to which the population is
homogeneous in its passive evolution provides a very sensitive measure
of the rate at which spheroidals may still be forming. Both tests are
valuable but progress on the latter has, until now, been hindered by
the small size of the spectroscopic samples available.

This paper is motivated by extending the early work begun in this area
via a comprehensive analysis of high quality Keck spectroscopy for 224
morphologically-selected spheroidals and bulge-dominated galaxies in
the northern field of the Great Observatories Origins Deep Survey
(GOODS). Associated with the Advance Camera for Surveys (ACS) and
spectroscopic data discussed here, is $K$-band photometry from which
stellar masses and optical-infrared colors have also been derived
elsewhere (Bundy, Ellis \& Conselice, 2005).

An overview of one of the main results of this observational program
has already been presented in Treu et al.\ (2005; hereafter T05),
where we discuss the evolution of dynamical mass-to-light ratios as a
function of mass and environmental density in the context of
hierarchical models of galaxy formation. We find that mass-to-light
ratios -- and hence luminosity weighted stellar ages -- depend
sensitively on the dynamical mass. The most massive galaxies have
evolved passively at least since $z\sim1.2$ while less massive systems
show strong evidence of recent star formation activity, consistent
with a ``downsizing'' scenario (Cowie et al.\ 1996). Within the range
of environments covered by the GOODS-N field, no dependency is seen
between this recent activity and the local galaxy density - contrary
to the expectations of hierarchical models of galaxy formation
(Diaferio et al.\ 2001; de Lucia et al.\ 2004).

In this paper, we considerably extend and quantify the arguments
presented in T05 in the context of a unifying hypothesis of the mass
assembly history of spheroidals of various masses. We describe
in detail our HST and Keck spectroscopic data and present the associated 
catalogs. We derive the evolution of mass-to-light ratios based
on analysis of the Fundamental Plane, taking into account
selection effects and correlating deviations from the overall trend
with morphological substructures.  

A plan follows. In $\S$2 we introduce the GOODS-ACS multicolor imaging
dataset, the construction of a photometric catalog and the procedure
whereby which morphological types were assigned to all galaxies
brighter than F850LP$<$22.5 (AB; hereafter $z_{\rm 9}$).  Surface
photometry and scale-lengths are fit to the sample of spheroidals and
morphological substructures identified via pixel-by-pixel color
imaging (c.f. Menanteau et al.\ 2001). $\S$3 presents the Keck
spectroscopy and its reduction and the techniques used to measure
stellar velocity dispersions. A complete catalog of the spectroscopic
and photometric data is presented. $\S$4 describes the correlation of
our catalog with the 2Ms X-ray catalog of the Chandra Deep Field North
(Alexander et al.\ 2003; Barger et al.\ 2003) to identify nuclear
activity. In $\S$5 we analyze the zero point, tilt, and scatter on the
fundamental plane as a function of redshift, and we use spectral and
photometric diagnostics to provide independent evidence for recent
star formation. $\S$6 summarizes our findings and discusses our
results in the context of formation scenarios for spheroidals.

We assume throughout a cosmological model with $\Omega_{\rm M}=0.3$, 
$\Omega_\Lambda=0.7$ and H$_0$=65 km~s$^{-1}$~Mpc$^{-1}$. All 
magnitudes are in the AB system (Oke 1974), unless otherwise noted.

\section{Imaging}

Imaging data are taken from the public GOODS survey (Giavalisco et
al.\ 2004). Initial sample selection and planning of spectroscopic
observations were performed before the completion of the survey and
are based on the v0.5 release of epoch 1 (as described in
\ref{ssec:photcat}). After completion of the GOODS-N survey and the
release of the combined v1.0 dataset, the photometric analysis was
repeated using the newly processed and deeper images as described in
this section.

\subsection{Photometric catalog}

\label{ssec:photcat}

In preparation for our spectroscopic observations in April 2003, a
first catalog was produced by running SExtractor (Bertin \& Arnouts
1996) on the v0.5 release of epoch 1 of the GOODS images. A $z_9$
selected catalog was constructed including all objects brighter than
MAG\_AUTO $z_{9,\rm v0.5}<22.5$. We will refer to this catalog as at
the v0.5-zcatalog. After the release of the combined GOODS images
(release v1.0) a further catalog was constructed using the same
procedure. We will refer to this catalog as the v1.0-zcatalog. The
v1.0-zcatalog is available from Bundy, Ellis \& Conselice (2005; URL
{\tt http://www.astro.caltech.edu/GOODS\_morphs/}).

Comparisons between the two catalogs provide an estimate of the photometric
accuracy. After adjusting for the different zero points (c.f. the
GOODS web page), the median difference and semi-interquartile range in
$z_{9}$ magnitudes between v0.5 and v1.0 are less than 0.01 and 0.03
mags respectively. Since these are negligible with respect to other
sources of errors throughout the paper, our catalog is effectively
magnitude selected in v1.0 as well. Accounting for the different zero
points, the selection limit corresponds to $z_{9,\rm v1.0}<22.43$.
Throughout the rest of the paper, all photometry is based on the v1.0
data and zero points.  Absolute magnitudes can be obtained from
apparent magnitudes and colors using the transformations described in
Appendix~\ref{app:restmag}.

\subsection{ACS Morphologies}

\label{ssec:morph}

In view of the spectroscopic run scheduled in March 2003, an initial
morphological classification was undertaken by one of us (RSE) by
visual inspection of the v0.5 epoch 1 $z_9$-band images.  The $z_9$
images were chosen to provide the closest match to optical rest frame
at a redshift $z\sim1$, minimizing the so-called morphological
$k$-correction. Morphologies (hereafter T) were assigned according to
the Medium Deep Survey scheme introduced by Abraham et al.\ (1996a;
see also Treu et al. 2003): T=-2=star, -1=compact, 0=E, 1=E/S0, 2=S0,
3=Sa+b, 4=S, 5=Sc+d, 6=Irr, 7=Unclass, 8=Merger, 9=Fault. When the the
deeper v1.0 data release became available, a blind re-classification
was also performed. Throughout the rest of the paper (unless otherwise
specified) and in the published catalog, all morphological types are
based on the deeper v1.0 data. Early-type and Sa+b galaxies in the
final spectroscopic sample are shown in Figures~\ref{fig:eviz1}
and~\ref{fig:sviz1}.

Comparison between the two morphological catalogs provides a check on
the internal consistency of the classification and on its sensitivity
to signal-to-noise (Figure~\ref{fig:morph}). Considering the entire
sample, the agreement is very good, with a median offset of 0, a mean
offset of -0.1 and a standard deviation of 1.3 (in the 12 point T
scheme described above).  However, for the class of spheroidal objects
with T=0,1,2 in the v0.5 catalog, the mean offset increases to 0.4
(median=0, standard deviation 1.2).  The differences arise because of
the improvement in signal-to-noise in the v1.0 images which reveal,
when present, the characteristic low surface brightness features of
spiral disks. Thus objects classed as spheroidals in the v0.5 catalog
(which formed the basis of our spectroscopic sample) include some
systems which later turned out to be bulge-dominated spirals, as
described quantitatively in Section~\ref{ssec:samplesel}.  Of course,
by including these outliers in our sample, we have a valuable measure
of the uncertainties associated with defining our morphological
boundaries.

Although we include our morphological classifications in the large
catalog discussed by Bundy, Ellis \& Conselice (2005), it is important
to examine the extent to which classification uncertainties permitted
in large statistical analyses might influence the results of this,
more focused, study of a relatively small number of spheroidals.  A
particular concern is the likelihood of contamination by the more
numerous bulge-dominated spirals.

We can estimate the probability of spiral contamination as follows.
At the magnitude limit of the survey, the rms scatter rms$_{\rm tot}$
in classification between the v0.5 and v1.0 catalogs is 1.3
types. Assuming that uncertainties scale as the square root of the
exposure time, we obtain rms$_{\rm tot}$=$\sqrt{\rm
rms_{1.0}^2+rms_{0.5}^2}$=$\sqrt{\rm rms_{1.0}^2+5 rms_{1.0}^2}$,
since the v1.0 images represent 5 times the integration of the v0.5
images.  This would appear to be a conservative assumption, since the
v1.0 images also have improved sampling, defect removal and relative
alignment. Thus, we expect rms$_{1.0}$=rms$_{0.5}/\sqrt(6)\sim0.5$
implying that $\sim$1/6 of the Sa+b sample might be misclassified as
S0s. As Sa+bs are twice as numerous as S0s overall, we estimate that
$\sim$1/3 of the S0s could be misclassified Sa+bs, i.e. approximately
20.

To identify possible misclassified Sa+bs, residuals from the \dv\,
fitting (faint analogs of the ones shown in
Figure~\ref{fig:galfitexamples}) were visually inspected by two of us
(RSE,TT) for disk asymmetries and spiral features.  To be
conservative, all edge-on S0s, where inclination could be hiding
spiral features, were also flagged. With these criteria, we identified
23 possible Sa+b contaminants (ID 199, 312, 369, 400, 402, 424, 505,
63, 655, 786, 811, 870, 872, 895, 996, 1286, 1441, 1463, 1485, 1491,
1577, 1685, 1764). The technique is surprisingly effective in locating
faint spiral features but we use this flagging only to illustrate the
effect of possible contamination rather than to refine the published
morphological catalog. We also identified 7 galaxies with prominent
dust lanes or asymmetries (ID 312, 377, 460, 879, 1354, 1477, 1764)
where the dust and asymmetries could bias photometric and
spectroscopic parameters used in our analysis. In
Section~\ref{ssec:FPquanti} we will test the robustness of our
analysis with respect by these possible biases by comparing results
obtained with and without these two subsamples.

\subsection{Surface photometry}

Surface photometry of all early-type galaxies and bulge dominated
spirals targeted for spectroscopy was performed using the publicly
available software GALFIT (Peng et al.\ 2002). Effective radius
(R$_{\rm e}$) and magnitudes (and hence effective surface brightness
SB$_{\rm e}$) were obtained in all four bands as the best fitting
parameters of a \dv\, surface brightness profile, following the
standard procedure used to construct the Fundmental Plane
(J{\o}rgensen, Franx \& Kj{\ae}rgaard 1992; van Dokkum \& Franx 1996;
Treu et al. 2001). The fitting procedure requires a point spread
function (PSF) to convolve the model before computing the
$\chi^2$. For each galaxy we adopted the nearest bright non-saturated
star in the GOODS-N field as fiducial PSF, thus correctly taking into
account the modifications of the PSF due to the GOODS v1.0 reduction
procedure. Examples of fits and residuals from a \dv profile are shown
in Figure~\ref{fig:galfitexamples}.

As is commonly known (e.g. Kelson et al. 2000a; Treu et al. 2001a) the
particular combination of photometric parameters necessary for
constructing the FP: FP$_{\rm ph}\,=\,\log R_{\rm e} - 0.32 SB_{\rm
e}$, is limited for datasets of good quality such as GOODS by the
fitting technique.  Tests with fit2D (Treu et al. 2001a) indicate that
the FP$_{\rm ph}$ parameter derived using GALFIT is consistent with
those derived in previous studies to a level of 0.02 or better. An
additional test can be obtained by comparing our photometry to the one
of vDE03, for the 9 objects in common. The effective radii agree
extremely well (average difference $\log R_{\rm e}$=0.0004, rms
scatter 0.066), and the photometry is in good agreement
(F814W-F775=0.010 mag, with rms=0.34 mag) considering the different
filters and instruments (assuming F814W=F775W). As a result FP$_{\rm
ph}$ agrees well (average difference 0.019, rms 0.051).

We adopt 0.03 as our best estimate of the systematic uncertainty on
the FP$_{\rm ph}$ due to the fitting technique.  Absolute magnitudes
can be obtained from apparent magnitudes and colors using the
transformations described in Appendix~\ref{app:restmag} (c.f. Treu et
al.\ 2001a). Rest frame effective radii are obtained by interpolation
from the two nearest filters as described in Treu et al.\ 2001a. 

\subsection{Internal colors, blue cores and morphological asymmetry}

\label{ssec:bcores}

The GOODS images are, like their HDF predecessors, of adequate signal
to noise for us to examine the internal homogeneity as a verification
of the morphological classifications and as a possible diagnostic of
recent activity. Of particular interest, given our large sample with
associated spectroscopy are the so-called `blue-core' spheroidals
first identified by Menanteau et al.\ (2001) in the HDF images. In this
paper we define a more robust way to uniformly select such systems at
various redshifts and our resulting sample is significantly larger
than any hitherto.

Rest frame pixel-by-pixel color maps (B-V) were derived by applying
the transformations described in Appendix~\ref{app:restmag}, with the
aim of identifying internal color peculiarities which might indicate
recent starformation activity. The vast majority of early-type
galaxies proved to be relatively smooth and homogenous.  However
14/163 early-type galaxies with measured redshift show a clear blue
core. 

Menanteau et al.\ (2001) already addressed the key issue of whether
the blue cores could be artefacts arising from the wavelength
dependency of the HST PSF. Although that analysis was conducted on the
basis of WFPC-2 images in the Hubble Deep Field and rich clusters, the
improved spatial sampling of the ACS makes their arguments tighter in
this case. An instrumental origin would imply a much greater frequency
of blue cores at a given apparent magnitude and not produce the
environmental dependencies discussed by Menanteau et al.\ or the
correlations with independent diagnostics of star formation discussed
below.

We chose to define a blue core spheroidal as one where a difference of
at least $\delta$(B-V)=0.2 is seen between the core of the galaxy and
the outer parts. As our color selection is made in rest-frame color,
this is significantly more useful than the observed $V-I$ criterion
used in previous studies (Menanteau et al. 2001, 2004, 2005).  We find
a total of 14 blue-core early-type galaxies in GOODS-N and these are
sorted by redshift and shown in Figure~\ref{fig:bcores}.  The nature
of these systems is discussed further in Sections~\ref{ssec:X}
and~\ref{ssec:FPadd}.

\section{Keck Spectroscopy}

\subsection{Sample selection and mask design}

\label{ssec:samplesel}

The sample for spectroscopic follow-up was selected from the
v0.5-zcatalog according to the following criteria: $z_{9,v0.5}<22.5$,
$-1\leq {\rm T}\leq 2$. Self-consistent astrometry across the field
was obtained by matching the ACS catalog to the ground based catalog
used by the Team Keck Redshift Survey (TKRS; Wirth et al.\ 2004).
Three sets of masks with their major axis aligned with that of the
GOODS-N mosaic were designed (north, center, south). For each pointing
a first submask was filled with objects at random from the sample of
early-type galaxies using the v0.5 morphologies. A second submask was
obtained by replacing objects brighter than $z_{9,v0.5}=21.5$ with new
objects. The goal of this procedure was to obtain longer exposure
times for the faintest objects while maximizing the number of survey
objects.

In total, 261 galaxies from the primary spectroscopic sample were
included in the six masks, plus 22 T=3 (v0.5) galaxies. Based on the
subsequently-determined deeper v1.0 morphologies, the spectroscopic
sample 175 early-types, 66 early-spirals (T=3), 14 later type galaxies
(T$>$4) and 28 stars. This mixing of types is expected because of the
increased signal-to-noise in the v1.0 (see section~\ref{ssec:morph})
images and because the number of objects that can drift out of the
sample due to random errors in the classification is much larger than
the number of objects that can enter, and because a conservative
star/galaxy separation criterion was used in the v0.5 images.

Eighteen slitlets (with a random distribution of morphologies and
luminosities) were not milled or could not be extracted because they
fell too close to the edges of the masks, so the final tally is
comprised of 265 objects. Six objects were observed twice, in
different slitlets of different masks, to allow for internal
checks. All the galaxies with measured redshifts are shown in
Figure~\ref{fig:eviz1} and~Figure~\ref{fig:sviz1}.

To ensure that the retrospective v1.0 morphological cut does not bias
the properties of our sample we compare in Figure~\ref{fig:sampling}
the luminosity and color distribution of the parent sample comprising
all early-type galaxies in the GOODS-N field (v1.0, empty histogram)
with defined in our spectroscopic sample (v0.5, hatched histogram).
The color distributions are very similar with a typical sampling rate
between 50 and 70\%, showing that our spectroscopic sample is unbiased
in color. However, the sampling rate is a declining function of
apparent magnitude beyond $z_{9}\sim21.5$, as expected, because of our
mask-design strategy, the higher surface density and the higher
probability of morphological misclassification.

In the following we will consider the early-type galaxies as our
primary sample. The early-spiral sample (composed mostly of objects
classified as early-type in the v0.5 analysis) will be useful as an
ancillary in order to compare with results based on shallower imaging
data. A complete discussion of the properties and star formation
histories of the bulges will be given elsewhere.

\subsection{Observations and data reduction}

The field was observed using the Deep Imaging and Multi-Object
Spectrograph (DEIMOS) at the Keck-II telescope on the nights of April
1 to 5 2003. Conditions were excellent, with clear sky and seeing in
the range $0\farcs6$-$0\farcs8$ throughout the run. The 1200 grating
blazed at 7500\AA\, was centered at 8000 \AA\, providing a pixel scale
of $0\farcs1185\times0.33$\AA, and a typical wavelength coverage of
2600 \AA. The $1\farcs0$ wide slitlets yielded an instrumental
resolution of $\sigma_{s}\sim 20-30$ kms$^{-1}$, as determined from
arc lines and sky emission lines. Total exposure times ranged between
14,400 and 38,900 seconds as listed in Table~\ref{tab:early}.

The data were reduced using the IDL pipeline developed by the DEEP2
collaboration (Davis et al.\ 2003; Faber et al. 2005, in prep). After
extraction, the 1d spectra were assigned a redshift using the software
developed by the DEEP2 collaboration (described in Coil et al.\
2004). All spectra were examined independently by two of us (TT and
TXL) and a quality criterion was assigned (q=-1 stars, q=2 possible,
q=3 likely, q=4 certain). In those very few cases where the
redshift/quality parameter differed, a simultaneous inspection by the
two classifiers was sufficient to reach an agreement.

Redshifts were obtained for all the 265 targeted objects, including
217 from the primary sample, 21 Sa+b galaxies, and 27 that turned out
to be stars (v0.5 classification). Based on the superior v1.0
classifications, the sample with redshifts is comprised of 163
early-type galaxies, 61 early-spirals, 23 later type galaxies and 1
star (according reclassified as compact).  Comparison with other
surveys shows that our redshifts are accurate to 30 kms$^{-1}$ or
better (c.f. Wirth et al. 2004).  An approximate relative flux
calibration was obtained using the measured sensitivity curves for
DEIMOS. Examples of spectra covering the range of redshifts and
luminosities (and hence signal to noise ratio) are given in
Figure~\ref{fig:specexamples}. The redshift distribution of the E+S0
and Sa+b samples is shown in Figure~\ref{fig:Nz}.

\subsection{Internal kinematics}

Stellar velocity dispersion were obtained by comparing the extracted
spectra with spectra of G-K giants and of the Sun (a high resolution
solar spectrum was obtained from the URL {\tt
http://bass2000.obspm.fr/solar\_spect.php}) as described in detail in
Treu et al.\ (1999; 2001) and in Sand et al.\ (2004). A brief summary
follows.

First, the instrumental resolution was measured by fitting gaussians
to well-exposed and unblended sky lines and arc calibration
lamps. This corresponds to a velocity dispersion of $\sigma_{\rm
s}=30-20$ \kms in the spectral range of interest, and is well
described by a second order polynomial in wavelength $\sigma_{\rm
s}(\lambda)$.  Second, high resolution template stars were redshifted
and smoothed to match the resolution of the instrumental setup, and
convolved with gaussians in $\log \lambda$ space to reproduce the
kinematic broadening.  The best fitting velocity dispersion was found
using the pixel fitting code developed by van der Marel
(1994). Spectral regions affected by emission lines, Balmer lines,
atmospheric A and B band absorption, as well as residual defects were
carefullly masked out during the fit.  Those pixels affected by the
strongest sky emission lines and their immediate neighbors were
similarly flagged. Each object was fitted independently with each
template and using different spectral regions.  The G-K giant library
extends from 4000 \AA\, redwards, while the solar spectrum extends to
the UV atmospheric limit, enabling us to take advantage of near-UV
absorption features in the highest redshift objects (c.f. van Dokkum
\& Stanford 2003).

Each spectrum and fitting result was carefully inspected by eye by one
of us (TT) to identify and correct potential problems (such as
insufficient wavelength range or reduction/extraction defects). Two
example of spectra with kinematic fits -- an intermediate quality
spectrum and the highest redshift object with measured velocity
dispersion -- are shown in Figure~\ref{fig:t15i7196}.

The majority (181/224 of all galaxies; 141/163 spheroidals) of objects
yielded a stable velocity dispersion measure ($\sigma_{\rm ap}$)
listed in Table~\ref{tab:early}. A few objects failed to deliver a
reliable velocity dispersion, due to insufficient signal to noise or
unfortunate location of the strongest absorption features on
persistent sky subtraction residuals. For a handful of objects at low
redshift the procedure yielded velocity dispersions below the
instrumental resolution $\sigma_{\rm s}=30$ \kms. These measures were
considered insufficiently reliable enough and discarded (although
plausibly correct given their faint intrinsic luminosity and the fact
that the NaD line is resolved). The success rate as a function of
redshift, luminosity, and average signal to noise ratio (per \AA,
observer frame) is summarized in Figure~\ref{fig:scomplete}.  For
simplicity, we report the average signal to noise ratio across the
entire spectrum as a measure of the total information, which is
approximately given by $<$S/N$>$ multiplied by the square root of the
number of \AA, i.e. $\sqrt{2600}$. The $<$S/N$>$ per \AA\, in the rest
frame is readily obtained as $<$S/N$>_{\rm rest}$=$<$S/N$>\sqrt{1+z}$.
As other authors (e.g. J{\o}rgensen, Franx \& Kjaergaard 1995) often
quote values based on smaller wavelength regions, or ones longward of
the 4000 \AA\ break, in comparing the limiting $S/N$ for which
dispersions were successfully measured, differences in definition
should be borne in mind (see Figure~\ref{fig:acfrsigma}).

For the FP analysis, the central velocity dispersions (i.e. within a
circular aperture of radius $r_{\rm e}/8$) was obtained as
$\sigma$=1.10$\pm$0.04 $\sigma_{\rm ap}$ (Treu et al.\ 2001).

\subsection{Tests of the Measured Velocity Dispersions} 

Given the size of our sample, a key issue before the scientific analysis 
that follows is the reliability of our stellar velocity dispersions. We have
approached this via several independent tests which, collectively,
demonstrate a high degree of precision has been attained, thanks
largely to the long integrations and the excellent quality of the data
obtained with the DEIMOS spectrograph.

\begin{enumerate}

\item {} Repeated measures of six objects observed with DEIMOS in different
slitlets/configurations (Figure~\ref{fig:acfrsigma}) show that our
measurements are stable with respect to internal effects such as
varying the wavelength range and the slitlet position in the field of
view of the spectrograph. 

\item {} A comparison with the 7 objects in common with van Dokkum \&
Ellis (2003) examined independently with the LRIS spectrograph shows
that both sets are also statistically indistinguishable, with $<\delta
\sigma / \sigma>=-0.06$ and an rms scatter of 12 \%, consistent with
the estimated errors ($0.12/\sqrt(6)=0.05$).

\item{} A subsample of spectra were analyzed independently by one of
us (PGD) using methods described in van Dokkum \& Franx (1996). The
internal comparison indicates stellar velocity dispersions that agree
to better than 1\% on average, with an rms scatter of 12\%, indicating
that there are no significant systematic uncertainties due to the
choice of fitting method.

\end{enumerate}

\noindent
We conclude that our estimated errors on velocity dispersion (on
average less than 10\%) are accurate.

\subsection{Basic Star Formation Diagnostics: [\ion{O}{2}] and H$\delta$}

\label{ssec:oiihd}

Whenever the redshift placed the useful diagnostic features of
[\ion{O}{2}] 3727 \AA\ and H$\delta$ 4101 \AA\ in the observed
spectral region, we measured their equivalent widths using an IDL
procedure.  The band definitions of Fisher et al.\ (1998;
[\ion{O}{2}]) and Trager et al. (1998; H$\delta_{\rm A}$) were
used. The noise properties of the spectra were taken into account so
as to maximize the signal-to-noise ratio. Observed frame equivalent
widths in \AA\, are listed in Table~\ref{tab:early} In the following
analysis we will always refer to rest frame equivalent widths in \AA.

\section{X-ray properties}

\label{ssec:X}

To obtain a complementary source of information on nuclear activity
and star formation in our sample of spheroidals (and in particular on
the blue cores) we cross correlated our redshift catalog with the
published X-ray 503 sources catalog of the Chandra Deep Field North
(Alexander et al. 2003; Barger et al. 2003), finding 42 sources with
X--ray counterparts (26 E+S0s; 13 Sa+bs; 3 Ss).  All the 42 sources
have redshifts published in Barger et al. (2003), in agreement with
our determinations.

From the soft and hard fluxes measured in Barger et al. (2003), we
derive the rest frame soft (0.5--2 keV) and hard (2--8 keV) emitted
luminosities applying a $k$-correction for a spectrum with photon
index $\Gamma=2$. A spectral slope of $\Gamma=2$ is appropriate for
X--ray emission from prompt star formation. Since the majority of the
42 X--ray counterparts of our spheroidal galaxies have emitted
luminosities in the range typical of star forming galaxies ($L_{0.5-2}
\leq 10^{42}$ erg s$^{-1}$, see Ranalli et al. 2003), we concluded
that $\Gamma=2$ is a good approximation of the required
$k$-correction. Adopting a different $k$--correction has a negligible
impact on our results, as described in the next paragraph.

For simplicity, we will classify our objects as AGN-dominated if the
X-ray luminosity (soft or hard) is above 10$^{42}$ erg s$^{-1}$, and
as star formation dominated or low luminosity AGN (LLAGN) if it is not
in the previous category and the X-ray luminosity is above 10$^{40}$
erg s$^{-1}$ (in the following we will label these two groups as XAGN
and XSF/LLAGN; motivations for these working definitions is given in
Appendix~\ref{app:X-ray}).  With these definitions, 17 sources (9
E+S0s) are XAGNs while 21 (16 E+S0s) are XSF/LLAGNs. Adopting
different slopes for the $k$-correction, e.g. $\Gamma \simeq 1.4$
typical of the average spectrum of absorbed and unabsorbed AGN that
dominate the population in the faint flux regime (as in the CFDN),
would only adjust these numbers in a minor way to 16 and 20 galaxies
respectively in the XAGN and XSF-LLAGN categories.  The same
classification is obtained by adopting the spectral slope evaluated
for each source from the X-ray hardness ratio (see Alexander et
al. 2003). Interestingly, only 2 of the 14 blue cores E+S0s are
classified as XAGN (ID=1323,872 Figure~\ref{fig:bcores}), while 1 is
classified as XSF-LLAGN (ID=1362), showing that most blue cores are
not due to a central AGN.

For the X-ray sources classified as XSF/LLAGN we can obtain an upper
limit to the star formation rate using the conversion $ \dot M = 2.2
\times 10^{-40} L_{0.5-2 keV}$ M$_\odot$ yr$^{-1}$ (Ranalli, Comastri
\& Setti 2003). The upper limits are in the range 1-100 M$_{\odot}$
yr$^{-1}$ with an average (median) value of 38$\pm$30 M$_{\odot}$
yr$^{-1}$ (28$\pm$23 M$_{\odot}$ yr$^{-1}$). We would like to
emphasize that in many of these cases, these upper limits will be well
above the actual star formation rate, because of the LLAGN
contribution and/or emission from X-ray halos and LMXB. This is
confirmed also by the optical spectra which do not show the dramatic
features that would be expected for star formation rates of order
$\sim $20 M$_{\odot}$ yr$^{-1}$. In order to disentangle more
accurately the various contributions -- especially the soft extended
halos from the typically hard and point-like AGN -- a detailed
analysis of the spectrum and the spatial distribution of the X-ray
emission will be necessary. This goes beyond the scope of the present
paper and is left for future work.

\section{Results}

\label{sec:res}

In this section we use our measurements to study the evolution of the
stellar populations of early-type galaxies, using the FP as a primary
diagnostic tool. In \S~\ref{sec:FPintro} we derive the formalism to
intrepret the evolution of slopes, intercept and scatter of the FP in
terms of (mass dependent) evolution of the effective mass-to-light
ratio. In the following sections, \S~\ref{ssec:FPquali} and
\S~\ref{ssec:FPquanti}, we will use the FP to study, first
qualitatively and then quantitatively, the evolution of the
mass-to-light ratio of spheroidal galaxies, showing that both the
slope and scatter of the FP change with redshift.  We will interpret
these results in terms of the evolution of the stellar populations. In
\S~\ref{ssec:FPcompa} we will compare our results with the existing
literature, showing that previously considered discrepant results can
be reconciled if the proper selection effects and uncertainties are
taken into account. Finally, in \S~\ref{ssec:FPadd}, we will examine
the relationship between the evolution of the mass-to-light ratio and
other stellar population diagnostics, providing independent evidence
for recent starformation.

\subsection{The Fundamental Plane as a diagnostic of stellar populations}
\label{sec:FPintro}

The Fundamental Plane is defined as
\begin{equation}
\label{eq:FP} \log R_{\tx{e}} = \alpha \log~\sigma + \beta~\tx{SB}_{\tx{e}} +
\gamma, 
\end{equation} 
($\sigma$ in \kms, R$_{\rm e}$ in kpc, SB$_{\rm e}$ in mag
arcsec$^{-2}$). For this study we will adopt as the local relationship
the FP of the Coma cluster ($\alpha=1.25$, $\beta=0.32$,
$\gamma=-8.970$, in B(AB) for H$_0=65$\kms Mpc$^{-1}$; J{\o}rgensen,
Franx \& Kj{\ae}rgaard 1996, hereafter JFK96).  We use Coma as the
local reference for both cluster and field to minimize systematic
uncertainties related to filter transformations, distance
determination, and selection effects; see discussion in Treu et al.\
(2001b). However, the effect of the systematic uncertainty in the
local intercept will be discussed and quantified in the analysis.  We
will also adopt 0.08 in $\log R_{\rm e}$ as the intrinsic scatter of
the local relationship, consistent with cluster and field estimates
(JFK96, Bernardi et al.\ 2003).

A simple physical interpretation of the FP (Faber et al.\ 1987) can be
given by defining an effective mass,
\be 
M \equiv \frac{5\sigma^2 R_{\tx{e}}}{G}, 
\ee
(c.f. Bender, Burstein \& Faber 1992), and by defining the luminosity
in the usual way
\be
-2.5 \log L \equiv \tx{SB}_{\tx{e}} - 5 \log R_{\tx{e}} - 2.5 \log 2 \pi,
\ee
and by considering an effective mass-luminosity relation of the form
(also known as the ``tilt'' of the FP; e.g., Ciotti, Lanzoni \&
Renzini 1996; Lanzoni et al.\ 2004)
\be
L \propto M^{\eta}
\label{eq:tilt}.
\ee
These assumptions lead directly to the FP relation given above,
provided
\be 
\alpha-10\beta+2=0, 
\label{eq:int}
\ee
(van Albada, Bertin \& Stiavelli 1995), with $\eta=0.2 \alpha /\beta$,
which is very closed to the observed value. When converted in solar
units and in the form of equation~\ref{eq:tilt}, reads:
\be
\log M/L_B = 0.25 \log M - 1.934.
\label{eq:massFP}
\ee

In this framework, variations of the slopes $\alpha$, $\beta$ and the
intercept $\gamma$ as a function of redshift are easily interpreted as
general trends in luminosity evolution of the stellar populations.  If
$\sigma$ and R$_{\rm e}$ do not evolve with redshift, for an
individual galaxy (labeled by the superscript $i$)
\be 
\gamma^i\equiv \log R_{\tx{e}}^i - \alpha \log \sigma^i - \beta \tx{SB}_{\tx{e}}^i,
\label{eq:gammai}
\ee 
the offset with respect to the prediction of the FP ($\Delta
\gamma^i \equiv \gamma^i-\gamma$) is related to the offset of the $M/L$ by
\be
\Delta \log \left( \frac {M}{L} \right)^i = -\frac{\Delta \gamma^i}{2.5 \beta},
\label{eq:dgdml}
\ee
which can be used to measure the average evolution and/or scatter of
$M/L$ at given $M$ (Note that equations~\ref{eq:gammai} and
\ref{eq:dgdml} are independent of the evolution of slopes and that
this relation does not require the FP to be of the form in
Equation~\ref{eq:massFP}.).  Therefore, the tightness of the FP
(JFK96; Bernardi et al.\ 2003) constrains the homogeneity of the
stellar populations of early-type galaxies (similarly to the small
scatter of the color-magnitude relation, Bower, Lucey \& Ellis
1992). An evolution with redshifts of the slopes of FP can be
interpreted in terms of a mass dependent evolution of $M/L$. For a FP
of the form in Equation~\ref{eq:massFP} the evolutionary rate depends
only on effective mass:

\be 
\frac{d \log (M/L) }{dz} = \frac{1}{2.5}\left[\frac{d\alpha}{dz}
\frac{\gamma-\log M - \log G + \log 5}{10\beta}-\frac{d
\gamma}{dz} \right].
\label{eq:massev}
\ee

In the following, we will often compare the evolution of the FP for
field early-type galaxies to that found for cluster E+S0s. For this
purpose, we will adopt as our fiducial evolution for E+S0 in massive
clusters
\be 
d \log (M/L_{\rm B})/dz= -0.46\pm0.04
\label{eq:vds} 
\ee
(van Dokkum \& Stanford 2003; see also van Dokkum et al. 1998, who
find $-0.49\pm0.05$). We also define a parameter that measures the
differential evolution between the mass to light ratio of a field E+S0
and that expected for the fiducial cluster E+S0:
\be 
\delta\Delta \log (M/L_{\rm B}) = \Delta \log (M/L_{\rm B}) + 0.46 z.
\label{eq:deltadelta}
\ee

\subsection{Evolution of the FP: qualitative results}
\label{ssec:FPquali}

The location of the E/S0 (pentagons) and Sa+b (stars) galaxies in the
FP-space is shown in the five panels of Figure~\ref{fig:FPev5Bn}, each
panel showing galaxies in a different redshift bin. The local
relationship measured in the Coma cluster (JFK96) is shown as a solid
line for reference (see Section~\ref{sec:FPintro} for discussion).
The early-type galaxies define a relatively narrow FP in each redshift
bin (note that the finite size of the bin introduces some artifical
scatter), while the scatter is larger for Sa+b galaxies. As
illustrated in T05, galaxies move away from the local relationship as
redshift increases, becoming brighter at fixed effective radius and
velocity dispersion.

The offset from the local relationship can be converted into a change
in effective mass-to-light ratio using Eqn~\ref{eq:gammai}
and~\ref{eq:dgdml}. The change for each individual galaxy as a
function of redshift is shown in Figure~\ref{fig:FPBpJ}. Evolutionary
tracks for a simple (single-burst) stellar population
\footnote{Unless otherwise stated all evolutionary models are computed
using Bruzual \& Charlot 1996 stellar population synthesis models
GISSEL96, Salpeter IMF, solar metallicity and Kurucz atmosphere
models.} formed at $z_f=1,2,5$ are also shown for comparison. It is
clear that even the genuine E+S0 population spans a wide range in
$\Delta \gamma_i$, covering the entire range predicted by these simple
models. We show these models only for illustration at this point; such
formation epochs would refer approximately to the luminosity weighted
age of the stellar population. Even a modest fraction of younger stars
would have a significant influence on the mass-to-light ratio (Trager
et al.\ 2000; Treu et al.\ 2001b).

As discussed in T05, the observed change in the slope in
Figure~\ref{fig:FPev5Bn} and increased scatter in
Figure~\ref{fig:FPBpJ} with respect to local samples, are the result
of mass-dependent trends, combined with selection effects and
increased intrinsic scatter of the FP. This is illustrated in
Figure~\ref{fig:FPBpJ} by plotting the offset from the local FP as a
function of galaxy mass. The most massive galaxies
(M$>10^{11.5}$M$_{\odot}$) have relatively high mass-to-light ratios,
minimal scatter and follow the passively-evolving track with a high
formation epoch fairly closely . As less massive galaxies are
included, the scatter and the average offset from the local relation
increase.  This possible trend was already tentatively seen in
previous studies (Treu et al.\ 2002; van der Wel et al.\ 2004; see
Pahre 1998, Kelson et al.\ 2000c, Wuyts et al. 2004, Moran et al.\
2004 for similar results in the cluster context), but only with our
large sample does it seem conclusive (T05). In the next section we
will obtain quantitative limits on the evolution of the mass-to-light
ratio and its scatter as function of mass and redshift taking due
account of selection biases.

\subsection{Evolution of the FP: quantitative results}
\label{ssec:FPquanti}

As we have seen in the previous section, the evolution of the FP
depends on galaxy mass, therefore we need to allow the FP slopes to vary
with redshift. In addition, since at face value it appears that the
intrinsic scatter around the FP increases with redshift, the question
of possible evolution in the intrinsic scatter is of great interest. As shown
in Treu et al.\ (2001b; 2002) the evolution of intercept, slope and
scatter are interconnected and deeply dependent on selection
effects. The larger the intrinsic scatter, the larger the Malmquist-like bias 
on the evolution of the mass-to-light ratio, and therefore the larger the 
correction necessary to recover the ``unbiased'' evolutionary trend.

The bias due to luminosity selection is illustrated in
Figure~\ref{fig:FPev5Bk} where we show the location of the early-type
galaxies in a different projection of the FP: the $M-M/L$ plane
(analogous to the $\kappa$ space introduced by Bender, Burstein \&
Faber 1992). The dotted line represents the local relationship, while
as in Figure~\ref{fig:FPev5Bn} the pentagons represent the early-type
galaxies. Hatched regions represent those excluded by our magnitude
limit.  As expected, objects with low mass and high mass to
light-ratio are excluded {\it a priori} from the sample. Only the two
lowest redshift bins probe deep enough to sample the entire FP. At
$z\sim0.5$ and above, selection effects begin to dominate the low mass
trends. Although we now embark on a rigorous modeling of the biases,
we can already infer from the existence of galaxies brighter than the
selection limit that the intrinsic scatter probably increases with
redshift and therefore that the smaller mass galaxies have a diverse
star formation history.

We note, in passing, that it is not sufficient to deduce evolutionary
trends by considering the high-mass objects only (e.g.  van der Wel et
al. 2004), since the uncertainties on M/L and M are
correlated. Selecting objects above a certain mass would introduce an
additional bias, leading in general to an overestimate of the M/L and
hence an underestimate of the evolution of this subset of the
population. In short, there is no way to avoid a proper modeling of
the observables.

\subsubsection{A Bayesan-Montecarlo approach}
\label{ssec:bayes}

Selection effects can be properly taken into account by using the
Bayesan-Montecarlo method introduced by Treu (2001) and Treu et al.\
(2001, 2002). Briefly, the method uses a Montecarlo algorithm to
compute the likelihood of obtaining the observed surface brightnesses
for fixed velocity dispersion and effective radius, given the
uncertainties and sample selection properties as a function of the FP
coefficients and scatter. The likelihood is then combined with a
prior using Bayes Theorem (1768) to recover the posterior probability
for each parameter. The likelihood is generally non-zero over a small
region of the possible parameter space and therefore the precise
choice of the prior is not critical, assuming that is chosen to be
regular enough. In the following, we adopt a uniform prior (see Treu
et al.\ 2001 for a discussion). All results presented in this refer
only to the spectroscopic sample of E+S0 galaxies, that is spirals are
excluded.

In this section we will derive formal limits on the evolution of the
coefficients of the FP using increasingly complicated models. As a
first step, we consider a simplified model where only the FP intercept
$\gamma$ and scatter are allowed to vary linearly with redshift, and
we thus seek limits on the derivative. This is the traditional
approach taken by most studies and therefore it is useful for
comparison with previous work. This procedure gives an estimate of the
average mass to light ratio evolution of the sample and an upper limit to
the evolution of the scatter. By construction the scatter for fixed
slopes is always larger than the scatter around the best fitting
slopes.  The posterior probability contours for our sample in this
case are shown in Figure~\ref{fig:Monte1}.

As expected the constraints on the two quantities are correlated in
the sense that less evolution in the mass to light ratio is required
for a population with larger scatter. The hypothesis that the FP
evolves with a fixed slope and constant scatter is rejected by our
dataset.  The best estimate and 68\% limits (obtained as described in
T01) are $d \gamma / dz=0.58^{+0.04}_{-0.06}$ and $d \sigma_{\gamma} /
dz = 0.18^{+0.03}_{-0.04}$ corresponding to $<d \log (M/L_{\rm B})/dz
>=-0.72^{+0.07}_{-0.05}$ and $d \sigma_{\log M/L} / dz<0.26$.  As
shown in Figure~\ref{fig:Monte1}, the average evolution of the
mass-to-light ratio for our sample of field E+S0s is significantly
faster than that of cluster E+S0s. 

To verify that our results are not influenced by the inclusion of
galaxies with dust-lanes or possible early-type spirals, we re-ran our
Montecarlo simulations excluding the objects flagged in
Section~\ref{ssec:morph}. Excluding the dusty/peculiar objects (5 with
measured velocity dispersion), changes the posterior probability in a
negligible way. Excluding the possible Sa+b galaxies (22 with measured
velocity dispersion) changes the above limits to $<d \log (M/L_{\rm
B})/dz >=-0.75^{+0.10}_{-0.05}$ and $d \sigma_{\log M/L} / dz<0.28$,
i.e. much less than the uncertainties (see Figure~\ref{fig:Monte1}).
We conclude that uncertainties in the morphological classification,
although likely at the level discussed, do not affect our results.

In addition, we compute the effects of systematic uncertainties on the
local intercept. For simplicity, we will express variations in the
local intercept as variations in surface brightness SB$_{\rm e}$ at a
fixed velocity dispersion and effective radius (c.f. Bernardi et al.\
2005). Since the evolution depends linearly on the local intercept,
there is a simple linear relation between uncertainties, i.e. $\delta
d log (M/L{\rm _B}) / dz$ = -0.57 $\delta$SB$_{\rm e}$, where the
coefficient is a weighted average of individual redshifts and offsets
(the coefficient has been verified numerically by repeating the
Montecarlo simulations after varying the local intercept). The
environmental dependence of the intercept in the local universe is
very small and hard to quantify because of selection effects, filters
and parameters definitions. As a conservative approach, we will adopt
the maximum environmental difference observed in the Sloan data
(Bernardi et al. 2005), corresponding to $\delta$SB$_{\rm e}=0.075$
mags, i.e. a systematic error on the evolutionary rate of $\pm0.04$.

Next, we consider a more realistic evolutionary scenario: namely
evolution of the slopes $\alpha$ and $\beta$, and hence mass dependent
evolution (Equation~\ref{eq:massev}). In implementing this model, we
assume that the FP can be described with a simple power law relation
between M and M/L at any redshift, i.e. $\alpha(z)-10\beta(z)+2=0$. We
will also assume that evolution of the coefficients and slopes can be
described by a linear relation,
e.g. $\alpha(z)=\alpha(0)+d\alpha/dz|_0\,z$. Thus our model has three
free parameters, $d\alpha/dz|_0$, $d\gamma/dz|_0$ and
$d\sigma_{\gamma}\, /\,dz|_0$. For simplicity, we will drop the index
0, assuming all derivatives are computed at $z=0$.

Posterior probability contours (68\% and 95\%) for $d \alpha / dz$ and
$d\gamma/dz$, marginalized over $d\sigma_\gamma/dz$, are shown in
Figure~\ref{fig:Monte2}. Constant slopes are clearly rejected,
demonstrating unambiguously that the tilt of the FP evolves with
redshift. This result is translated in terms of effective mass to
light and mass in Figure~\ref{fig:Monte3} using
Eqn~\ref{eq:massev}. The evolution of the mass to light ratio depends
strongly on mass: the most massive galaxies evolving similarly to
massive cluster galaxies (i.e. $d \log (M/L_{\rm B})/dz=-0.46$), while
evolution is faster for smaller masses. This is the {\it downsizing}
trend discussed in T05.

As a final step, we derive the posterior probability distribution
function for the evolution of the intrinsic scatter of the FP
$d\sigma_{\gamma}/dz$ -- marginalized over $d\alpha/dz$ and
$d\gamma/dz$ -- shown in Figure~\ref{fig:Monte4}. The peak value of
the probability distribution of function and the 68\% contours are
$d\sigma_\gamma/dz=0.032\pm0.012$, corresponding to $d{\rm rms}(\log
M/L_{\rm B})/dz=0.040\pm0.015$. Note this measurement of the scatter
is much smaller than the scatter at fixed slopes
(Figure~\ref{fig:Monte1}), showing that part of the visual impression
of the scatter when the evolution is plotted as in
Figure~\ref{fig:FPBpJ} is dominated by the change in the slope.  This
proves that the scatter in the FP, and therefore the scatter in
stellar population properties, increases with redshift, as expected
for example if the scatter is due to a spread in stellar ages.

\subsection{Implications for the star formation history of E+S0 galaxies}
\label{ssec:sfh}

In this section we proceed to interpret our findings on the evolution
of the FP in terms of star formation and assembly history of E+S0
galaxies. Since only objects morphologically identified as E+S0 are
included in our sample, what we describe is the effective star
formation history of the existing population at any given redshift. If
there was significant morphological evolution (``progenitor bias'', as
discussed by van Dokkum \& Franx 2001 for cluster early-type
galaxies), the average star formation activity of the parent
population would be more delayed.

First, we will use single burst stellar population models showing that
they imply very low redshift of formations for the population with
effective mass below 10$^{11}$ M$_{\odot}$, inconsistent with
independent star formation diagnostics and other observational
facts. Then, we will proceed to explore more realistic models of
composite stellar populations.

The traditional physical description of the evolution seen in the FP
is the change in mass to light ratio expected in the redshift range
$z=0-1.2$ for a simple (single-burst) stellar population (SSP) (the
correspondence between $d \log (M/L_{\rm B}) /dz$ and redshift of
formation $z_f$ is shown in Figure~\ref{fig:zf}).  For early-type
galaxies more massive than $10^{11} M_{\odot}$, the evolution implies
a high redshift of formation, $z_f>2$, whereas redshifts of formation
as low as $z_f=1.2-1.3$ are adequate for the systems of lower
effective mass.

A major result of our paper, however, is that the above description is
too simplistic because various independent diagnostics available to us
(see Section~\ref{ssec:FPadd}) indicate that a significant fraction of
the early-type population with mass lower than 10$^{11}$ M$_{\odot}$
has suffered recent/secondary star formation. Furthermore, a SSP
redshift of formation of $z_f\sim1.2$ for lower mass objects would be
impossible to reconcile with the results of deep infrared surveys
(Glazebrook et al.\ 2004; Cimatti et al.\ 2004), which have found a
substantial number of galaxies (including morphological E+S0; e.g.,
Stanford et al.\ 2004) with stellar mass above 10$^{10}$ M$_{\odot}$
at redshift above 1.2.

A more appropriate representation is therefore a composite stellar
population. Since subsequent bursts involving a relatively small
fraction of stellar mass can substantially alter the evolution of the
mass to light ratio (Trager et al.\ 2000; Treu et al.\ 2001), we will
consider models where a (dominant) fraction of the stellar mass is old
($z_{f1}>2$) and the rest of the stellar mass was formed at
$z_{f2}<1.2$. We will show that these models provide a good description
of the data and use them to derive limits on the fraction of younger
stars.

This procedure is illustrated in Figure~\ref{fig:zf2} where we compare
the evolution of the mass-to-light ratio for families of simple
two-burst models with that observed for our sample. Starting from the
left panel panel we discuss E+S0 galaxies in order of decreasing mass.
For the most massive galaxies ($>10^{11.5} M_{\odot}$), the average
evolution is well described by that of a very old stellar population
formed at $z_f=5$. Secondary burst at $z_{f2}<1.2$ are allowed but
they cannot involve more than 1\% of the total stellar mass. For
intermediate mass objects (center panel), larger bursts are allowed,
up to $\sim 5$ \% of the stellar mass. For the smallest masses (right
panel) bursts involving 10\% of the stellar mass at $z<1.2$ are
needed. For the largest masses these are solid upper limits and are
consistent with, for example, accretion of small satellites. As
discussed in T05 this is at variance with the expectations of a simple
scheme where the growth of stellar mass follows the bottom-up growth
of dark matter halos (see T05 for discussion).

For the less massive galaxies, however, when the mass in the secondary
bursts starts to be a significant fraction of the total mass there is
an additional complication. These larger secondary bursts are a quite
dramatic event, which could be the result for example of a major
merger, involving a rapid transformation of gas into stars. After such
a dramatic event, it is unlikely that a galaxy would be recognizable
morphologically until dynamical relaxation is complete (i.e. of order
a Gyr after the event; c.f. van Dokkum \& Franx 2001). Thus a more
realistic estimate of the mass in these significant secondary bursts
is obtained as in T05, assuming at least 1Gyr of age for the young
population. In some cases, this requires fraction of mass in the young
population as high as 20-40\%.

\subsection{Comparison with previous results}
\label{ssec:FPcompa}

In the past few years, several groups have used the FP to investigate
the evolution of the mass-to-light ratio of field spheroidal galaxies
either using direct measurements of velocity dispersion (Treu et
al. 1999, 2001a,b, 2002, 2005; van Dokkum et al.\ 2001; van Dokkum \&
Ellis 2003; Bernardi et al.\ 2003; Gebhardt et al.\ 2003; van der Wel
et al.\ 2004) or estimating stellar velocity dispersions from the
separation of multiple images of gravitationally lensed background
objects (Kochanek et al.\ 2000; Rusin et al.\ 2003; van de Ven et al.\
2003; Rusin \& Kochanek 2005). The Lenses Structure and Dynamics (LSD)
Survey (Koopmans \& Treu 2002,2003; Treu \& Koopmans 2002,2003,2004;
hereafter collectively KT) measured stellar velocity dispersions of
gravitational lenses thus bridging the gap between the two methods.
For reference, previous measurements are summarized in
Table~\ref{tab:previous}.

The first thing to notice from the table is that all published
measurements indicate faster FP evolution than that found for cluster
galaxies ($-0.49\pm0.05$, van Dokkum et al.\ 1998) although in some
cases the difference is not statistically significant. The second
important point is that most measurements, even those based on lensing
properties, agree with one another to within the quoted 1-$\sigma$
uncertainties.  This can be explained by the structural homogeneity of
early-type galaxies which appear to have close to isothermal mass
density profiles (e.g. Rusin et al. 2003b) as a result of a still
unexplained dark- luminous matter conspiracy (e.g. Treu \& Koopmans
2004).

If we take the measurement from this paper as the reference, all
measurements listed in Table~\ref{tab:previous} are in agreement
within the uncertainties except for van Dokkum \& Ellis (2003) and
Rusin et al.\ (2003) who measure a slightly slower evolution, and
Gebhardt et al.\ (2003), whose work appears to indicate faster
evolution. The latter discrepancy might be explained by a selection
bias of the type shown in Figure~\ref{fig:FPev5Bk}, particularly since
the Gebhardt et al study used considerably shorter exposure times at
comparable redshifts.  Other differences might be explained by the
different mass ranges in the various samples, contamination by
bulge-dominated spirals in shallower imaging data, or simply by small
number statistics and different fitting techniques. If mass-to-light
ratios follow the tracks illustrated in Figure~\ref{fig:zf2}, as a
result of a series of secondary bursts overlayed on an old stellar
population, the average evolutionary rate will depend on the fraction
of objects caught in the active luminous phase. Similarly, different
fitting techniques could be more or less sensitive to ``outliers'' and
yield varied answers and error estimates.  In short, small samples
will be very sensitive to fluctuations in this number and to the
adopted fitting technique.

Previous studies have often emphasized the differences between the
results, arguing whether or not field early-type galaxies have younger
stellar populations than their counterparts in clusters. Our sample
allows us to firmly establish that field early-type galaxies have, on
average, younger stellar populations than cluster galaxies, as a
result of a more diverse star formation history, i.e. a wider range in
epochs of star formation and secondary activity extended at relatively
recent cosmic times.

\subsection{Independent diagnostics of recent star formation}
\label{ssec:FPadd}

In T05 we showed that the mass-to-light ratio measured from the FP
correlates well with the rest frame B-V color, indicating that
deviations from the passive trend (defined in that paper by that
observed for cluster early-types; and measured with the parameter
$\delta \Delta M/L_{\rm B}$, Section~\ref{sec:FPintro}) is associated
with younger stellar populations. In this Section, we will extend this
analysis, using spectral ([\ion{O}{2}] and H$\delta$) and
morphological (blue cores) diagnostics as independent probes of the
star formation history of our sample of E+S0 galaxies. We will use
[\ion{O}{2}] as an indicator of ongoing star formation (e.g. Kennicutt
1992), since the life time of the hot OB stars required to produce
photo-ionizing photons is very short. Similarly, as strong H$\delta$
absorption occurs primarily in A stars with main sequence lifetimes of
$\sim$1 Gyr, we use its occurrence to measure the recent activity.

Prior to using blue cores and emission lines as star formation
diagnostics, we must consider what fraction of this activity might be
due to non-thermal AGN. For this purpose we identify nuclear activity
on the basis of X-ray luminosity as discussed in Section~\ref{ssec:X}.
Adopting a conservative threshold of $L_X\,>\,10^{42}$ erg s$^{-1}$,
we find only two of the 14 galaxies with blue cores are likely to be
AGN dominant, with a further a possible LLAGN.  Similarly, three
strong [\ion{O}{2}] emitters (i.e. EW$>15$\AA; 2 early-types, 1 Sa+b)
without blue cores are identified as AGNs. These objects are excluded
from the analysis in the rest of this section, guaranteeing that the
remaining blue cores and [\ion{O}{2}] are indeed due to star
formation.

To visualize the connection between star formation indicators,
morphology and mass to light ratio, we plot in Figure~\ref{fig:DelBC}
the distribution of the offset from the redshift-dependence of the FP
evolution of massive cluster spheroidals $\delta \Delta M/L_{\rm B}$
(Equation~\ref{eq:deltadelta}). Galaxies with signs of recent star
formation (blue-cores, strong [\ion{O}{2}] emitters, and H$\delta$
equivalent widths$ <-3$\AA ) show in general a larger offset than the
quiescent population. However, even the quiescent population evolves
more rapidly than the cluster population ($\delta \Delta M/L_{\rm
B}\equiv0$). Quantitatively, for the E+S0 galaxies, the average
$\delta \Delta M/L_{\rm B}$ is $-0.23\pm0.02$ for the entire sample,
$-0.39\pm0.07$ for the H$\delta$ strong galaxies, $-0.39\pm0.10$ for
the star forming blue cores, and $-0.17\pm0.12$ for the star forming
[\ion{O}{2}] emitters.

In conclusion, galaxies whose images and spectra reveal evidence for
recent star formation have lower mass to light ratios at a given
redshift than their quiescent counterparts. No significant difference
is found for the sample of galaxies with [\ion{O}{2}] emission,
perhaps because of small number statistics (11 E+S0s, excluding the 2
AGN) or perhaps because of the different timescales ([\ion{O}{2}] is a
measure of instantaneous star formation, while $M/L_{\rm B}$ is a
measure of integrated star formation).  However, the fraction of
[\ion{O}{2}] E/S0 emitters (11/54) for which the line is visible with
our spectroscopic setup is significantly higher than for early-type
galaxies in the local field ($\sim 0.05$ Treu et al.\ 2002), providing
further evidence for enhanced star formation.

Studying the evolution of H$\delta$ absorption we can obtain an
independent check on our inferences based on the evolution of the mass
to light ratio and perform an independent test of the secondary
activity postulated in Section~\ref{ssec:bayes}. In
Figure~\ref{fig:Hd2} we show H$\delta$ equivalent width vs. redshift
for all E+S0 galaxies \footnote{Note that H$\delta$ is only visible
for a subset of objects (74/163 E+S0s), i.e. when redshifted into the
wavelength range covered by our spectrograph.}  and compare it with
the predictions of our simple (thick red line) and composite
population models (black dashed lines). We find that the distribution
of rest-frame H$\delta$ equivalent widths is inconsistent with any
single burst model formed at high redshift and recent star formation
activity is required. As for the M/L$_{\rm B}$, the distribution is
well-reproduced by a composite population where the bulk of stellar
mass is formed at high redshift (e.g. $z_{f1}\sim3$) and on average 10
\% of the stellar mass is added at a later time ($z_{f2}\sim0.5-1$) in
secondary bursts.

Evidence for downsizing in the population, discussed initially by T05,
is also seen in the H$\delta$ absorption. To illustrate this, coadded
spectra of all the E+S0 galaxies binned in redshift and mass are shown
in Figure~\ref{fig:coaddev2}. For galaxies with masses smaller than
10$^{11}$ M$_{\odot}$ we find an average H$\delta$ equivalent width
$\sim -3$~\AA, inconsistent with the expected value for an old stellar
population (c.f. Figure~\ref{fig:coaddev2}; see also van Dokkum \&
Ellis 2003 calculate $\sim -1$\AA\, for stellar populations formed at
$z\sim3$ and observed at $z\sim0.9$).

\section{Summary and conclusions}

We have obtained ultra-deep spectroscopy of 163 field E+S0 and 61
bulge dominated Sa+b galaxies, selected morphologically from the
GOODS-N field to a magnitude limit of $z_{\rm9}<22.43$. Redshifts are
measured for all objects in the sample, while stellar velocity
dispersions are obtained for 141/163 E+S0s and 40/61 Sa+bs. The E+S0s
with measured stellar velocity dispersion represent an unbiased and
representative subsample of the parent morphological
population. Structural parameters are obtained by fitting \dv profiles
to the ACS images. Morphological asymmetries and blue cores are
identified on the basis of the multicolor ACS images. Blue core
spheroidals represent approximately 10\% of the population
(14/163). When visible with our spectroscopic setup, [\ion{O}{2}] and
H$\delta$ are measured and used as secondary star formation
indicators.

Our redshift catalog is cross-correlated with the deep 2Ms X-ray
catalog of the Chandra Deep Field North (Barger et al.\ 2003) to
identify AGN and star forming galaxies. Nine E+S0s, including 2 blue
cores, and 6 Sa+bs are identified as AGN based on an X-ray luminosity
above $10^{42}$ erg s$^{-1}$. 21 additional galaxies (16 E+S0s, 4
Sa+bs, 1S) are also detected in the X-ray with luminosities between
$10^{40}$ and $10^{42}$ erg s$^{-1}$, which are interpreted as mostly
due to a low luminosity AGN or to ongoing star formation.  The vast
majority (11/14) of blue core E+S0s are not detected in the deep
Chandra exposures, indicating that the blue cores are the result of
spatially concentrated star formation activity, and not of non thermal
AGN emission.

We study the evolution of the Fundamental Plane as a diagnostic of the
star formation and mass assembly history of early-type galaxies, using
Montecarlo simulations to rigorously control selection effects.  The
main results can be summarized as follows:

\begin{enumerate}

\item The mass-to-light ratio of field spheroidals evolves on average
as $d\log (M/L_B) /dz=-0.72^{+0.07}_{-0.05}\pm0.04$, i.e. faster then
that of cluster spheroidals ($-0.49\pm0.05$ van Dokkum et al.\
1998). This is consistent with the stellar population of field
spheroidals being on average younger. Although the difference between
the FP evolution of cluster and field spheroidals (Table 2) is in the
sense expected by hierarchical models (Diaferio et al. 2001; De Lucia
et al.\ 2004), quantitatively it is not as pronounced as expected (van
Dokkum et al.\ 2001; Treu 2004).

\item The evolution of the Fundamental Plane, and thus of the
effective mass to light ratio, depends strongly on mass: the less
massive the galaxy, the faster the evolution. The mass-to-light ratio
of the most massive galaxies ($>10^{11.5}M_{\odot}$) evolve as that of
an old stellar population ($z_f>2$), with an upper limit of $\sim 1\%$
to the fraction of stellar mass that can have formed at $z<1.2$. The
evolution of the mass to light ratio of the less massive galaxies is
inconsistent with that of a single burst stellar population. A
significant fraction of the mass (up to 20-40 \% for masses below
$10^{11}$ M$_{\odot}$) must have formed more recently than
$z_{f2}=1.2$ consistent with significant recent growth of this
population.

\item The intrinsic scatter of the FP is found to increase with
redshift, after allowing for evolution of the slopes, as $d {\rm rms}
(\log M/L_{\rm B}) /dz = 0.040\pm0.015$, consistent with the idea that
we have approached an epoch of starformation activity in this
population.

\item The scenario of mass dependent star formation history and
significant recent growth for E+S0 with mass below $10^{11}$
M$_{\odot}$ is supported by independent star formation indicators.
The mass-to-light ratio of spheroidal galaxies with blue rest frame
colors (discussed in a companion paper, T05), strong emission or
balmer absorption ([\ion{O}{2}] and H$\delta$), and morphological
anomalies (blue cores) is found to evolve more rapidly than that of
quiescent galaxies. The fraction of E+S0s galaxies with [\ion{O}{2}]
emission is larger (13/54, or 11/54 excluding two emitters identified
as AGN based on the X-ray emission) than in the local universe
($\sim$0.05; Treu et al.\ 2002 ). The distribution of H$\delta$
absorption for our sample is well reproduced by the same two-bursts
models describing the evolution of the FP.

\end{enumerate}

We promote, therefore, a picture in which all spheroidals except the
most massive ones continue to assemble and form young stars via
subsequent episodes of star formation. The associated rejuvenation is
sufficient to significantly influence the location of the galaxy on
the FP and, of course, its colors.  As a result, a fraction of
spheroidal galaxies will have blue colors, and be missed in color
selected samples (e.g. Jimenez et al.\ 1999).  Likewise, passive
models will be an inadequate representation of the evolutionary
behavior. Introducing secondary bursts implies that for a given
luminosity function (or equivalently observed surface density; Im et
al. 2002; Cross et al.\ 2004) will lead to an increased evolutionary
rate in the number density evolution (Treu 2004).

Ultimately we seek to connect the evolution of the mass to light ratio
for various masses with the broader issue of how the {\it number} of
spheroidals is growing with cosmic time (c.f. Franx 1993).  To
illustrate a possible approach to this important larger picture, we
explore combining the FP measurements made in this paper with
constraints on the evolution of the luminosity function (LF), for
example morphologically-selected E+S0s from the WFPC2 survey discussed
by Im et al (2002).

Im et al. (2002) chose to represent the evolution of their E+S0 LF
with two parameters: $Q$ describing the luminosity evolution
(equivalent to $-2.5 d \log (M/L_{\rm B}) / dz$), and $m$ describing
the number density evolution, i.e. the characteristic density
parameter in the Schechter (1976) luminosity function which is
postulated to evolve as $(1+z)^{m}$.  With luminosity function data
alone, there is a degeneracy betweem $Q$ and $m$: stronger luminosity
evolution implies a stronger number density evolution. Using our
measured value for the luminosity evolution inferred independently
from the FP (equivalent to $Q=1.80\pm0.15$) in combination with the
$\chi^2$ surface shown in their Figure~24, we eliminate the degeneracy
and obtain $m=-0.6\pm0.5$.

This value of $m$ implies an abundance of morphologically selected
E+S0s at $z\sim1.2$ of $60^{+30}_{-20}$ \% of the local value. If
correct, the decline in the number density of red galaxies (Bell et
al.\ 2004) would be explained in part by an evolution of the number
density of E+S0s galaxies (at least those morphologically
identifiable, e.g., van Dokkum \& Franx 2001) and in part by their
blue colors. Such an interpretation would apply equally to samples of
color- and morphologically- selected higher redshift E+S0 galaxies
(Menanteau et al. 1999; Treu \& Stiavelli 1999; Stanford et al. 2004).

Although a crude resolution of the differences seen for the various
datasets, largely because of the uncertainties in combining different
datasets and the limited size (and thus cosmic variance noise) of the
Im et al.\ 2002 sample, by combining of our precise FP measurements
with more extensive luminosity functions a more robust empirical
picture of the evolution of E+S0s in the last 9 Gyrs can be
determined.

\acknowledgments

The use of the Gauss-Hermite Pixel Fitting Software developed by
R.~P.~van der Marel is acknowledged.  Financial support by NASA
(Hubble Fellowship HF-01167.01; STScI-AR-09960) and NSF (AST-0307859)
is acknowledged. We thank Kevin Bundy for his help in developing the
morphological classification catalog and for numerous scientific
conversations. We thank those who developed DEIMOS and the staff of
the Keck Observatory for making this project possible and the DEEP2
team for encouragement and helpful discussion. In particular, we would
like to thank Greg Wirth, Chris Willmer and the Keck GOODS team for
kindly providing the self-consistent astrometric solution that was
used to design the slit-masks, and Sandy Faber and David Koo for
scientific discussions. We acknowledge helpful discussions with
B. Abraham, M. Bernardi, G.Bertin, K. Glazebrook, S. Gallagher,
X. Hernandez, S.Pellegrini, A. Renzini, S.White.  We are grateful to
Myunshin Im for providing his $\chi^2$ surface in digital format. This
paper is based on data obtained with the HST operated by AURA for NASA
and the W.M. Keck Observatory on Mauna Kea, Hawaii. The W.M. Keck
Observatory by the California Institute of Technology, the University
of California and NASA and was made possible by the generous financial
support of the W.M.  Keck Foundation.

\clearpage

\begin{deluxetable}{llllllllllllllllll}
\tablecaption{Catalog of spheroidals and bulge dominated galaxies \label{tab:early}}
\tabletypesize{\scriptsize}
\setlength{\tabcolsep}{0.06in}
\tablehead{
\colhead{ID} & \colhead{RA} & \colhead{DEC} & \colhead{T} & \colhead{b$_{\rm 4}$} & \colhead{\resecb} & \colhead{v$_{\rm 6}$} & \colhead{\resecv } & \colhead{i$_{\rm 8}$} & \colhead{\reseci } & \colhead{z$_{\rm 9}$} & \colhead{\resecz} & \colhead{$z$} & \colhead{$\sigma_{\rm ap}$} & \colhead{[\ion{O}{2}]} & \colhead{H$\delta$} & \colhead{S/N} & \colhead{t$_{\rm exp}$}}
\startdata
1287 & 189.13248 & 62.15182 & 0 & -     & -    & 21.90 & 0.81 & 20.80 & 0.45 & 19.95 & 0.47 & 0.8457 & 319$\pm$22 & -0.4$\pm$0.3 & 1.3$\pm$0.2  & 22 & 19800 \\
 648 & 189.16547 & 62.13849 & 0 & 22.30 & 0.76 & 21.21 & 0.37 & 20.58 & 0.29 & 20.24 & 0.28 & 0.2469 & 65$\pm$13  & -            & -            & 26 & 19130 \\
 609 & 189.19130 & 62.10799 & 0 & -     & -    & 22.45 & 0.74 & 21.09 & 0.55 & 20.51 & 0.49 & 0.7976 & 247$\pm$16 & 5.4$\pm$0.6  & -2.2$\pm$0.4 & 12 & 19130 \\
1286 & 189.13710 & 62.15260 & 2 & -     & -    & 22.41 & 1.21 & 21.06 & 0.85 & 20.36 & 0.76 & 0.8461 & 231$\pm$16 & -0.6$\pm$0.6 & 1.7$\pm$0.4  & 12 & 19800 \\
 655 & 189.19568 & 62.13973 & 1 & -     & -    & 21.91 & 0.78 & 20.72 & 0.81 & 20.35 & 0.75 & 0.6421 & 99$\pm$7   &  -           & 3.7$\pm$0.4  & 14 & 19130 \\
1253 & 189.13830 & 62.14290 & 3 & -     & -    & 22.23 & 0.03 & 21.90 & 0.03 & 21.24 & 0.10 & 0.9349 &        -   & 3.8$\pm$0.2  & 1.8$\pm$0.2  & 18 & 19800 \\
1236 & 189.13470 & 62.13730 & 0 & -     & -    & 22.91 & 0.61 & 21.59 & 0.40 & 20.92 & 0.35 & 0.8503 & 202$\pm$11 & 0.5$\pm$0.4  & -2.0$\pm$0.3 & 14 & 37800 \\
1278 & 189.15150 & 62.14850 & 2 & -     & -    & 22.51 & 0.14 & 21.57 & 0.12 & 21.19 & 0.12 & 0.5125 & 108$\pm$7  &  -           & -            & 22 & 37800 \\
 681 & 189.16202 & 62.15102 & 1 & -     & -    & 23.28 & 0.40 & 21.87 & 0.34 & 21.17 & 0.32 & 0.8422 & 318$\pm$28 & -0.7$\pm$0.5 & -0.3$\pm$0.3 & 13 & 38930 \\
 635 & 189.20297 & 62.12884 & 2 & -     & -    & 23.24 & 0.59 & 21.73 & 0.45 & 21.05 & 0.41 & 0.8195 & 187$\pm$16 & -0.5$\pm$0.5 & -2.1$\pm$0.4 & 11 & 38930 \\
\enddata

\tablecomments{For each object we list ID, RA, DEC, morphological
type, AB magnitudes (from \dv\, fitting) and effective radii (in
arcsec), redshifts, measured velocity dispersion $\sigma_{\rm ap}$ in
\kms, [\ion{O}{2}] and H$\delta$ equivalent widths (in \AA, rest
frame), average signal-to-noise ratio (\AA$^{-1}$; observer frame) and
spectroscopic exposure time (in seconds). Photometric parameters
through the $b_4$ filter are derived only for objects at $z<0.34$.
Formal galfit uncertainties on magnitudes and effective radii are
0.01-0.02 mags and a 1-10\% in effective radii. Total uncertainties on
the combination of photometric parameters that enters the Fundamental
Plane $\log R_{\rm e}-0.32 {\rm SB}_{\rm e}$ are 0.02-0.03. The Table
is published in its entirety online, the first 10 entries are shown
here for guidance.}
\end{deluxetable}

\begin{deluxetable}{lllll}
\tablecaption{Summary of previous field measurements \label{tab:previous}}
\tabletypesize{\small}
\tablehead{
\colhead{$d \log (M / L_{\rm B})/dz$} & \colhead{b$_{\rm Redshift}$} & \colhead{sample} & \colhead{Source} & \colhead{Notes}}
\startdata
-0.50$\pm$0.19          & 0.0-1.0  & 22     & Rusin \& Kochanek 2005   & lensing-general; lenses\\
-0.54$\pm$0.06          & 0.6-1.0  & 9      & van Dokkum \& Ellis 2003 & direct; morphology and magnitude\\
-0.56$\pm$0.04          & 0.0-1.0  & 27     & Rusin et al.\ 2003       & lensing-isothermal; lenses\\
-0.59$\pm$0.15          & 0.2-0.6  & 18     & van Dokkum et al.\ 2001  & direct; morphology and magnitude\\
-0.62$\pm$0.13          & 0.0-1.0  & 26     & van de ven et al.\ 2003  & lensing-isothermal; lenses\\
-0.71$\pm$0.19          & 0.6-1.1  & 6      & van der Wel et al.\ 2004 & direct;  morphology, magnitude and color\\
-0.72$^{+0.16}_{-0.11}$ & 0.1-0.7  & 30     & Treu et al.\ 2002        & direct; morphology, magnitude and color\\
-0.72$\pm$0.10          & 0.5-1.0  & 5      & Treu \& Koopmans 2004    & direct; lenses, early-types only \\
-0.75$\pm$0.17          & 0.5-1.0  & 5      & Treu \& Koopmans 2004    & lensing-isothermal; lenses, early-types only \\
-0.80..-96              & 0.3-1.0  & 36     & Gebhardt et al. 2003     & direct; magnitude and absorption lines\\
-0.72$^{+0.07}_{-0.05}\pm0.04$ & 0.2-1.2  & 141    & this work         & direct; morphology and magnitude\\
\enddata

\tablecomments{In the notes column, ``direct'' refers to
spectroscopically determined velocity dispersions,
``lensing-isothermal'' refers to velocity dispersion estimate from
image separation using an isothermal model, ``lensing-general'' refers
to velocity dispersion estimate from image separation using a more
general model (see Rusin \& Kochanek 2005); the measurements for
Gebhardt et al.\ 2003 are the average brightening to $z=0.8$ and
$z=1.0$, respectively.}
\end{deluxetable}

\clearpage

\appendix

\section{Transformations from ACS magnitude to rest frame B and V magnitudes}
\label{app:restmag}

Throughout this paper, rest frame Landolt B and V magnitudes are
obtained from the two ACS filters which are closest to the redshifted
B and V central wavelengths (i.e. 4405 and 5470 \AA, e.g., Fukugita et
al.\ 1996).  Given the redshift interval considered here $0<z<1.25$, a
pair of ACS filters is always sufficient to obtain rest frame B and V
magnitudes within few hundreds of magnitudes. Hence, the
transformations adopted here can be expressed in the form:

\begin{equation}
B=v_6+\alpha_{\rm Bvbv}(b_4-v_6)+\beta_{\rm Bvbv} - {\rm DM} \quad [0<=z<0.34]
\label{eq:Bvbv}
\end{equation}

\begin{equation}
B=i_8+\alpha_{\rm Bivi}(v_6-i_8)+\beta_{\rm Bivi} - {\rm DM} \quad [0.34<z<=0.76]
\label{eq:Bivi}
\end{equation}

\begin{equation}
B=z_9+\alpha_{\rm Bziz}(i_8-z_9)+\beta_{\rm Bziz} - {\rm DM} \quad [0.76<=z<1.25]
\label{eq:Bziz}
\end{equation}

\begin{equation}
V=v_6+\alpha_{\rm Vvbv}(b_4-v_6)+\beta_{\rm Vvbv} - {\rm DM} \quad [0<=z<0.07]
\label{eq:Vvbv}
\end{equation}

\begin{equation}
V=i_8+\alpha_{\rm Vivi}(v_6-i_8)+\beta_{\rm Vivi} - {\rm DM} \quad [0.07<z<=0.41]
\label{eq:Vivi}
\end{equation}

\begin{equation}
V=z_9+\alpha_{\rm Vziz}(i_8-z_9)+\beta_{\rm Vziz} - {\rm DM} \quad [0.41<=z<1.25]
\label{eq:Vziz}
\end{equation}

\noindent where the coefficients $\alpha$ and $\beta$ are a function
of redshift, while the bolometric distance modulus DM is a function of
redshift and cosmography. Polynomial expansions of the coefficients
are given in Table~\ref{tab:photocoeff}. 

\begin{table*}[h]
\begin{center}
\begin{tabular}{lllllll}
\hline
N     & $\alpha_{\rm Bvbv}$ & $\beta_{\rm Bvbv}$ & $\alpha_{\rm Bivi}$ & $\beta_{\rm Bivi}$ & $\alpha_{\rm Bziz}$ & $\beta_{\rm Bziz}$ \\
\hline
0     &  0.824411  & -0.00169895  &  2.44316  & 0.748236 &  1491.01 & -796.805 \\
1     & -3.29360   &     1.22223  & -4.42301  & -3.72483 & -8947.85 &  4776.19 \\
2     &  3.40645   &    -8.36258  &  1.53361  &  7.43317 &  22264.2 & -11813.0 \\
3     &        &     18.9313  &           & -3.58153 & -29340.5 &  15426.2 \\
4     &        &          &           &          &  21564.4 & -11200.7 \\
5     &        &          &           &          & -8367.60 &  4278.26 \\
6     &        &          &           &          &  1336.45 & -669.418 \\
\hline
\hline
N     & $\alpha_{\rm Vvbv}$ & $\beta_{\rm Vvbv}$ & $\alpha_{\rm Vivi}$ & $\beta_{\rm Vivi}$ & $\alpha_{\rm Vziz}$ & $\beta_{\rm Vziz}$ \\
\hline
0     & 0.186741 & -0.0274962 &  1.42588  & -0.0394309  & -77.0651  &  46.5707 \\
1     & -2.62562 &  0.995802  & -5.13394  &    1.23626  &  668.581  & -399.767 \\
2     &  7.19822 &  0.0925006 &  4.22521  &   -1.42416  & -2294.21  &  1398.85 \\
3     &          &            &           &    1.88833  &  4056.19  & -2539.31 \\
4     &          &            &           &             & -3936.67  &  2531.96 \\
5     &          &            &           &             &  1999.86  & -1319.01 \\
6     &          &            &           &             & -417.432  &  281.589 \\
\hline
\end{tabular}
\end{center}
\caption{Taylor expansion coefficients of $\alpha$ and $\beta$ around
redshift $z=0$. Redshift intervals of validity are given in
Equations~\ref{eq:Bvbv} to~\ref{eq:Vziz} \label{tab:photocoeff}}
\end{table*}

\clearpage

\section{X-ray-based classification scheme}
\label{app:X-ray}

In this paper we adopted a simplified classification scheme, based on
X-ray luminosity, to identify galaxies with nuclear activity (XAGN),
or with low luminosity nuclear activity and/or ongoing star formation
(XSF/LLAGN).  The working definitions of AGN or XSF/LLAGN --
admittedly schematic -- are motivated by local studies of early-type
galaxies which show that typically an AGN is required to power a
luminosity above 10$^{42}$ erg s$^{-1}$ (with the exception of cluster
brightest cluster galaxies -- not present in our sample -- where the
ICM can reach these luminosities e.g. Pellegrini \& Ciotti 1998,
O'Sullivan, Forbes \& Ponman 2001). At fluxes below 10$^{42}$ erg
s$^{-1}$ the main contributors to the X-ray luminosities are LLAGNs,
star formation related phenomena such as HMXB, low-mass X-ray binaries
(LMXB), and emission from the diffuse plasma, with tipical
temperatures below 1 keV.

We can estimate the contribution of the two latter mechanisms as
follows.  The contributions from LMXB is proportional to the stellar
mass and thus scales linearly with the B--band luminosity and it is
generally small with respect to the total X--ray emission observed in
local early--type galaxies (see O'Sullivan, Forbes \& Ponman 2001; Kim
\& Fabbiano 2004). Using the relation from Kim \& Fabbiano L$_X$
(LMXB) = (0.9$\pm$0.5) 10$^{30}$ erg s$^{-1}$ L$_{\rm B}$ / L$_{\rm
B,\odot}$ and our measured L$_{\rm B}$ we find that for our
sample\footnote{$\log L_{\rm B}/L_{\rm B,\odot}$ is in the range
9.2-10.8 average 10.2, after evolving to $z=0$ using the evolution of
the FP measured in Section~\ref{sec:res}.} the median contribution of
LMXBs to the total X--ray emission is 16$\pm$9\%.  

As far as diffuse plasma emission is concerned, we can conservatively
estimate its contribution by using the empirical scaling of the total
soft X-ray luminosity with L$_{\rm B}$ as measured for E+S0 galaxies
in the local universe, excluding AGN hosts and brightest cluster
galaxies. Since local early-type galaxies are mostly quiescent, the
excess with respect to the local relationship provides an estimate of
the contribution due to star formation or LLAGN in high redshift
sources.  For example O'Sullivan, Ponman \& Collins (2003) find
$L_{0.5-2} \simeq 4 \times 10^{41} \times (L_B/10^{11})^{2.7}$, albeit
with a scatter of an order of magnitude. Applied to our sample, this
relationship results in an expected X-ray luminosity corresponding to
6$\pm$4\% of the total observed one. As an example of the uncertainty
in this estimate, adopting instead the relationship in Eq.1 from
O'Sullivan, Forbes \& Ponman (2001) raises the fraction to
$25\pm15$\%. We conclude that in general the X-ray luminosity of our
distant E+S0 with luminosities in the range 10$^{40}$-10$^{42}$ erg
s$^{-1}$ is due largely to star formation and/or LLAGN and cannot be
explained entirely by the plasma and LMXB emission expected for local
E+S0. However, given the high intrisic scatter of the local
relationships, XSF/LLAGN might not be needed in some individual cases.

\clearpage

\begin{inlinefigure}
\begin{center}
\resizebox{\textwidth}{!}{\includegraphics{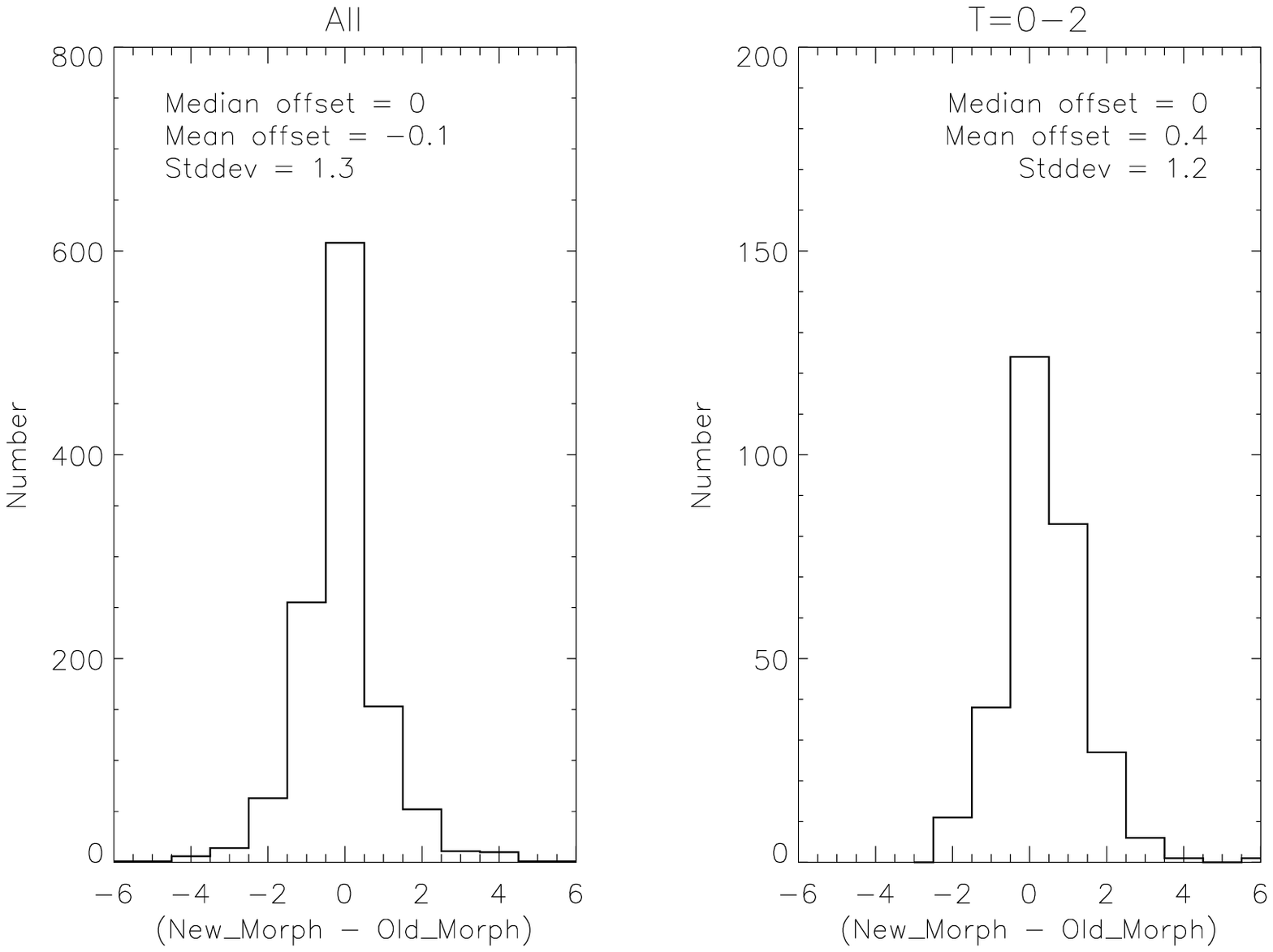}}
\end{center}
\figcaption{Comparison of morphological classifications based on the
v0.5 and v1.0 releases of the GOODS data, referred to as ``old'' and
``new'' respectively. Types T are assigned based on the following
scheme: T=-2=star, -1=compact, 0=E, 1=E/S0, 2=S0, 3=Sa+b, 4=S, 5=Sc+d,
6=Irr, 7=Unclass, 8=Merger, 9=Fault (Abraham et al. 1996; Treu et
al. 2003). Left panel refers to all classes, right to spheroidals
only.
\label{fig:morph}}
\end{inlinefigure}

\begin{figure*}
\begin{center}
\end{center}
\figcaption{Color-composite images (R=$z_{\rm 9}$, G=$i_{\rm 8}$,
B=$v_{\rm 6}$) of all early-type galaxies in the Keck spectroscopic
sample, sorted by redshift. The color scheme is chosen so that a local
early-type galaxy would appear white.  Images are $3''$ on a side. For
each galaxy redshift, ID, and morphological type are shown in the
upper left, lower left and lower right corner respectively.
\label{fig:eviz1}}
\end{figure*}

\begin{figure*}
\figurenum{2}
\leavevmode
\begin{center}
\end{center}
\figcaption{Color-composite images of all early-type galaxies (continued).
\label{fig:eviz2}}
\end{figure*}

\begin{figure*}[t]
\figurenum{2}
\leavevmode
\begin{center}
\end{center}
\figcaption{Color-composite images of all early-type galaxies (continued).
\label{fig:eviz3}}
\end{figure*}

\begin{figure*}[t]
\begin{center}
\end{center}
\figcaption{Color-composite images (R=$z_{\rm 9}$, G=$i_{\rm 8}$,
B=$v_{\rm 6}$) of all the early-spirals in the Keck spectroscopic sample,
sorted by redshift.
\label{fig:sviz1}}
\end{figure*}

\clearpage

\begin{figure*}[t]
\begin{center}
\resizebox{\textwidth}{!}{\includegraphics{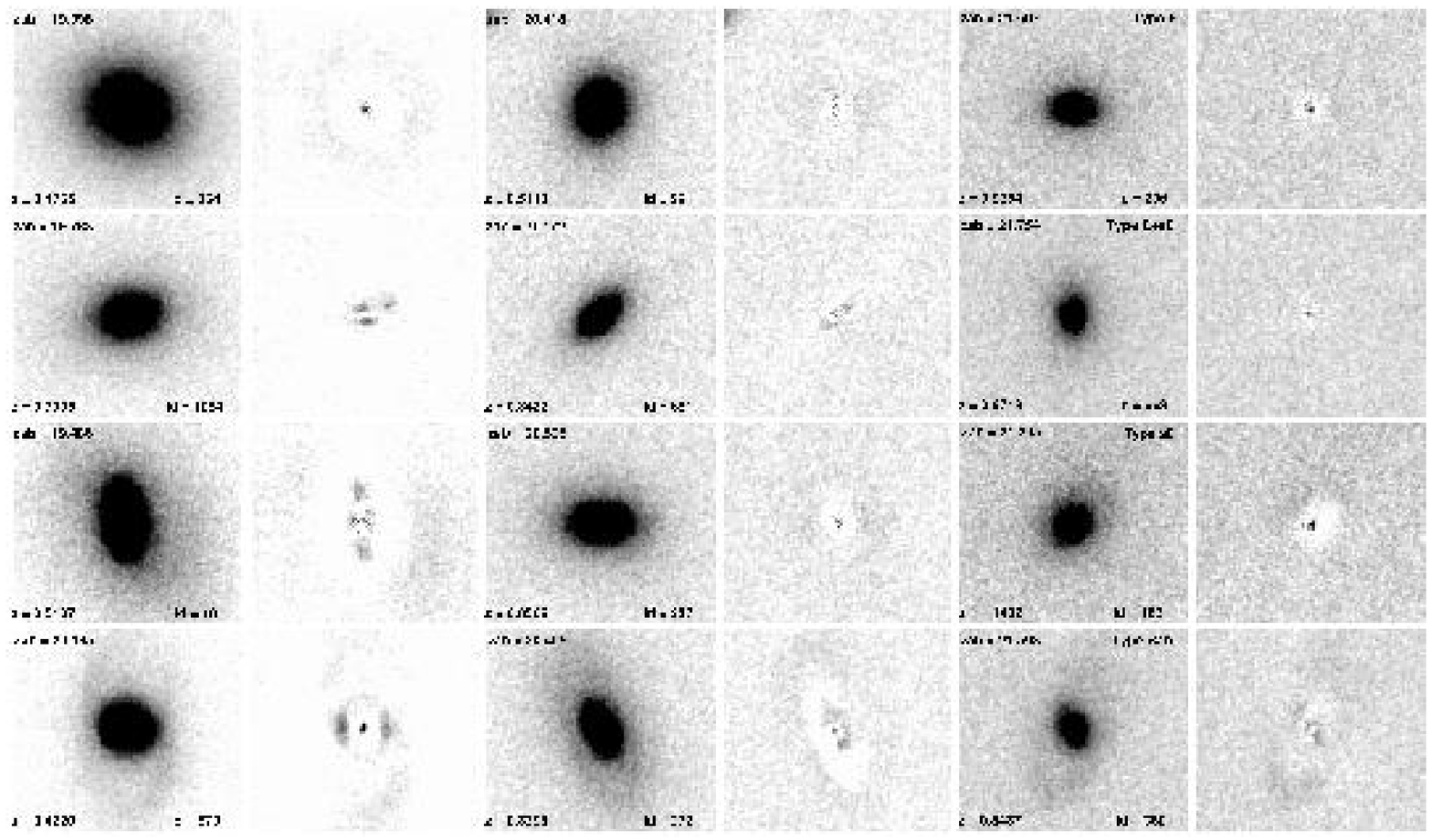}}
\end{center}
\figcaption{Examples of de Vaucouleurs fits: for galaxy types $-1<T<4$
(one per row), for three different magnitude bins, $z\sim20$,
$z\sim21$, $z\sim22$. For each galaxy the original image is shown alongside
residuals from the best fit \dv model. Notice the disk-like
residuals in the S0 galaxies and the spiral arm residuals in the Sa+b
galaxies confirming the v1.0 visual classification. Redshift,
magnitude and ID are listed for each object. Each panel is
$3\farcs0\times2\farcs7$.
\label{fig:galfitexamples}}
\end{figure*}

\begin{figure*}
\begin{center}
\end{center}
\figcaption{Montage of 'blue core' early type galaxies, selected as
defined in the text, sorted by redshift. Each image is 3'' on a
side. Only the two highest redshift examples appear to host active
galactic nuclei, as indicated by the large X-ray luminosity
($>10^{42}$ erg s$^{-1}$). The bulk of the blue light arises from
recent star formation.
\label{fig:bcores}}
\end{figure*}

\clearpage

\begin{inlinefigure}
\begin{center}
\resizebox{\textwidth}{!}{\includegraphics{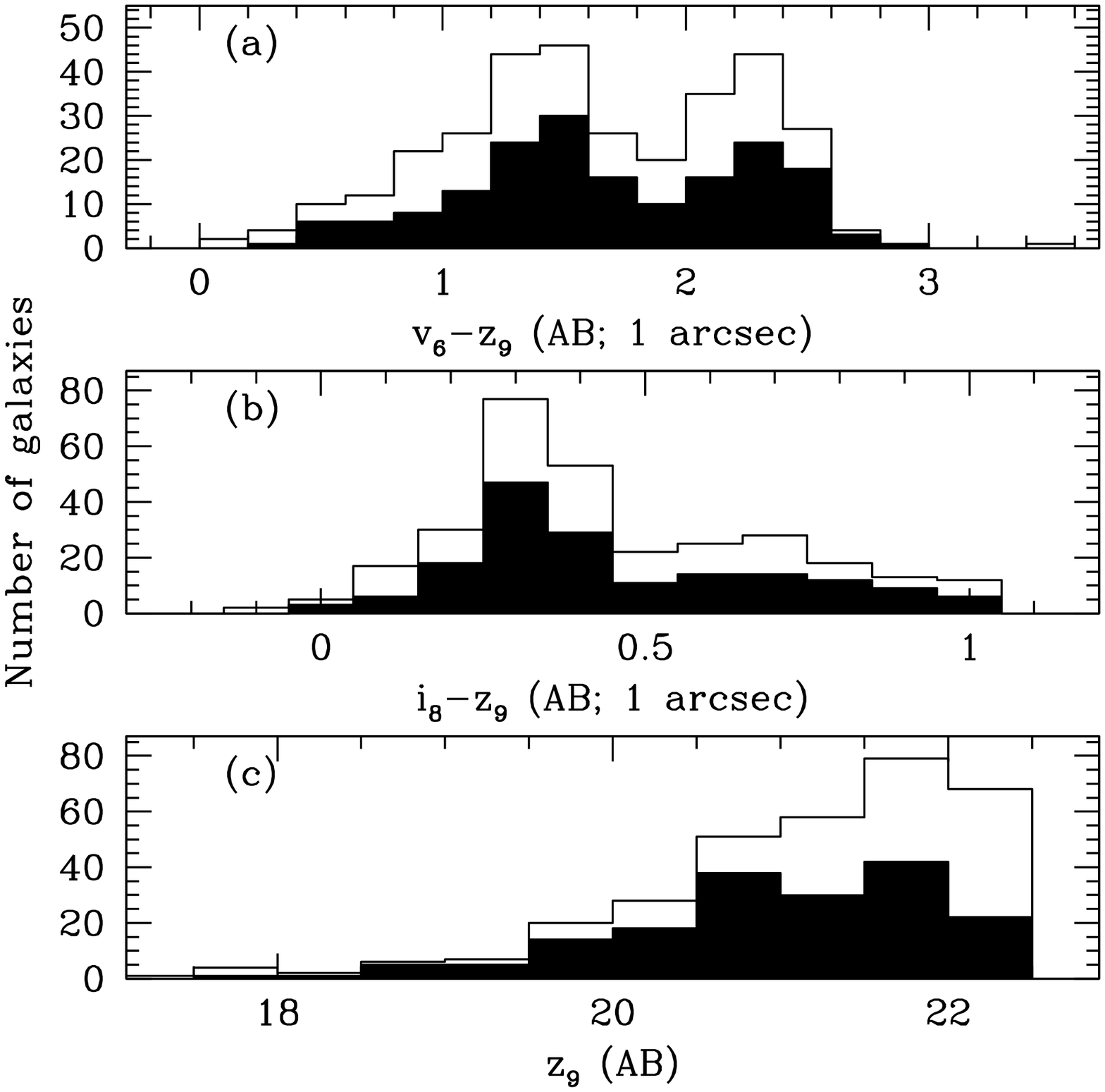}}
\end{center}
\figcaption{Distribution of colors (panels a and b) and $z_{\rm 9}$
magnitudes (panel c) for the full sample of $z_{\rm 9}<$22.43 E+S0 
galaxies classified in the final v1.0-zcatalog (empty histogram) compared 
to the E+S0 sample (v1.0 morphologies) targeted for spectroscopy 
(solid histogram). The sampling rate is approximately independent of 
color, but drops off beyond $z_{\rm 9}\sim21.5$ as discussed in 
Section~\ref{ssec:samplesel}.
\label{fig:sampling}}
\end{inlinefigure}

\begin{figure*}
\begin{center}
\resizebox{\textwidth}{!}{\includegraphics{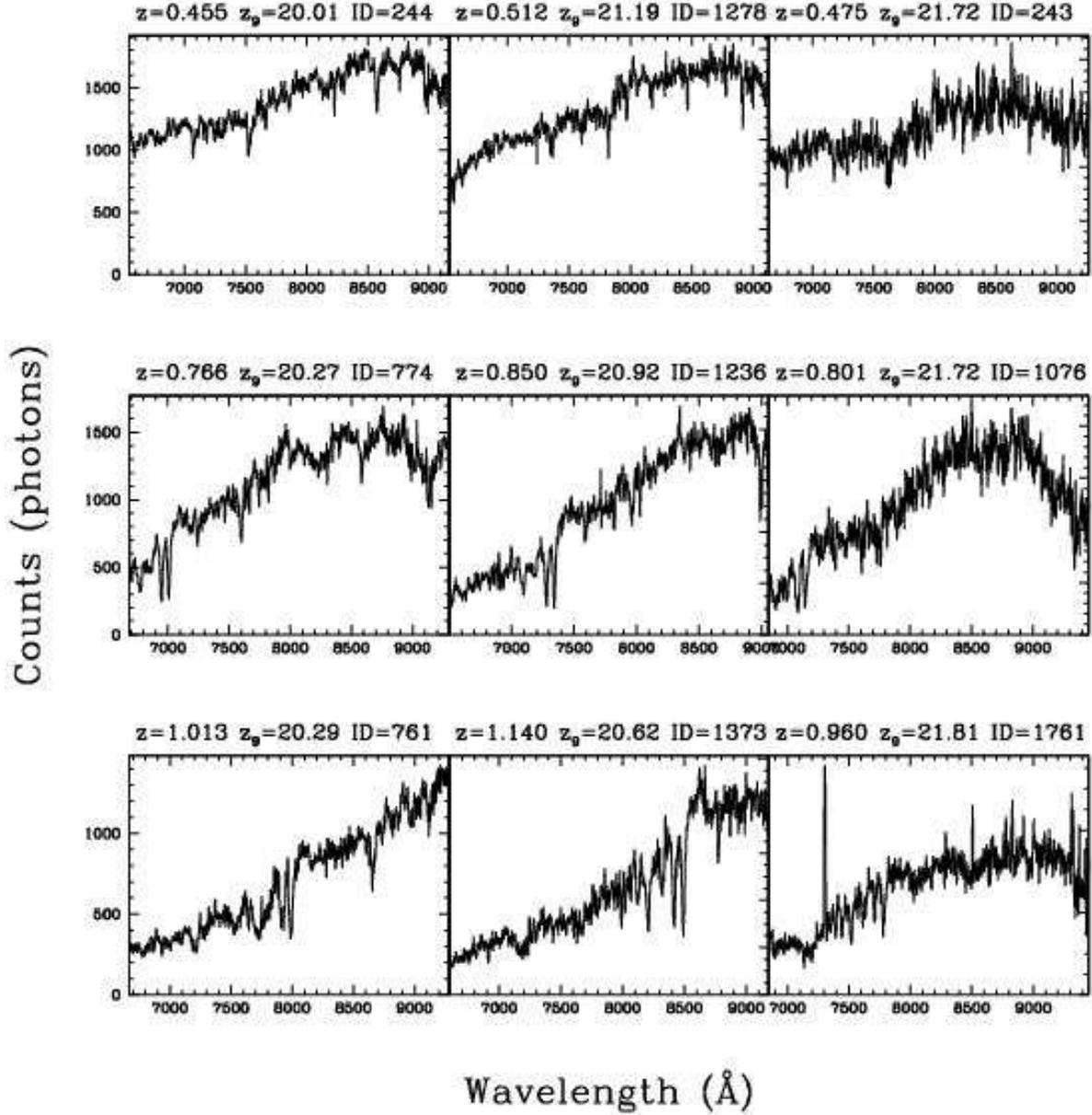}}
\end{center}
\figcaption{Examples of flux-calibrated spectra -- smoothed with a
boxcar filter of width 6\AA -- for three different magnitude bins,
$z_9\sim20$, $z_9\sim21$, $z_9\sim22$ and redshifts
($z\sim0.5$,$z\sim0.8$,$z\sim1$). The signal-to-noise ratio decreases
with magnitude and exposure time as expected. Fluxes are in units
proportional to photons per second.
\label{fig:specexamples}}
\end{figure*}

\begin{inlinefigure}
\begin{center}
\resizebox{\textwidth}{!}{\includegraphics{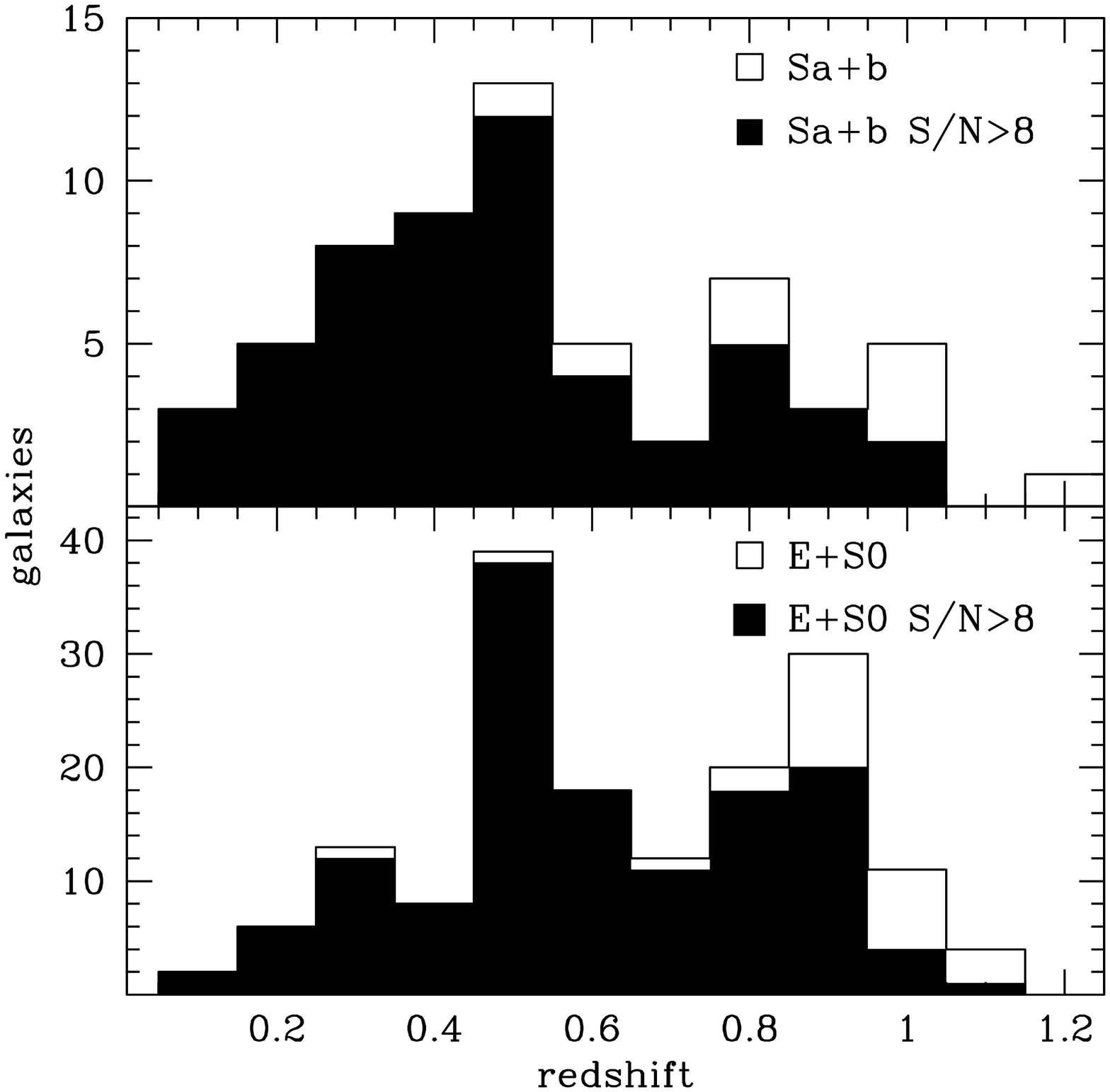}}
\end{center}
\figcaption{Redshift distribution of the early type spiral and E+S0
sample.  Spectra with S/N$>$8 \AA$^{-1}$\, in the observers frame, for
which velocity dispersions are typically available, are shown as a
filled histogram.
\label{fig:Nz}}
\end{inlinefigure}

\begin{inlinefigure}
\begin{center}
\resizebox{\textwidth}{!}{\includegraphics{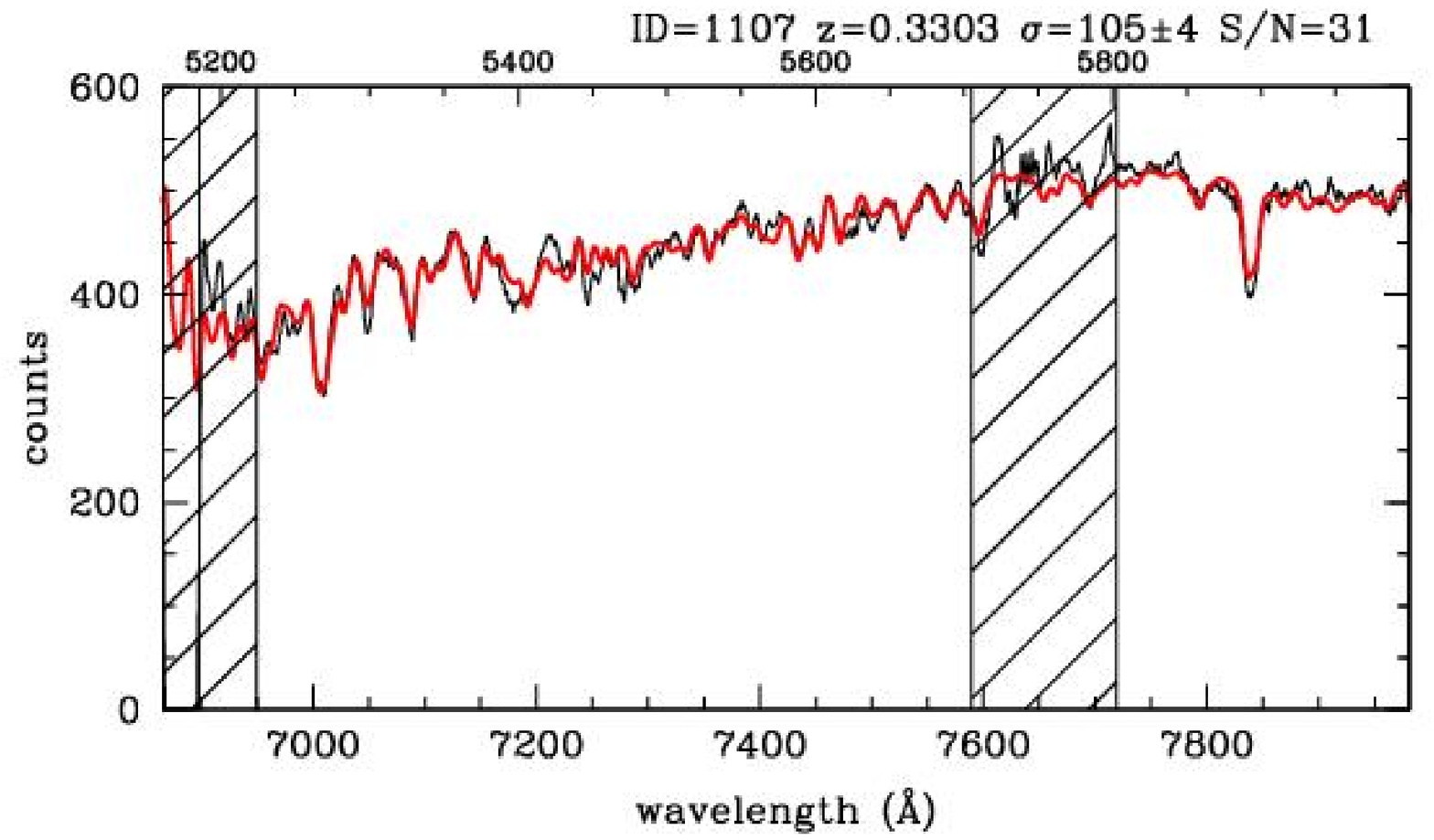}}
\resizebox{\textwidth}{!}{\includegraphics{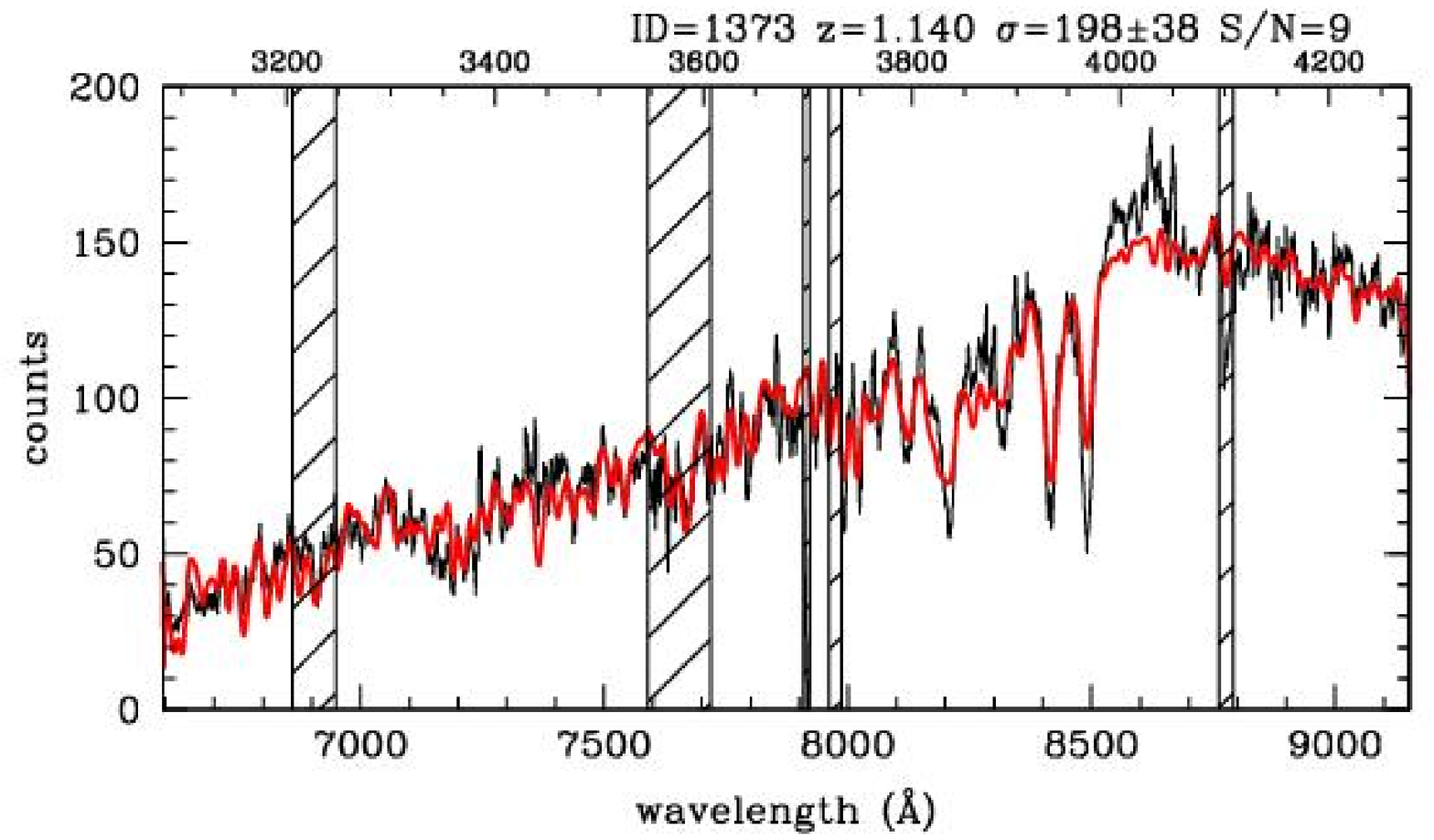}}
\end{center}
\figcaption{Example of an intermediate S/N spectrum (top) and the
highest redshift spectrum with measured velocity dispersion
(bottom). Both spectra are boxcar smoothed with a 6\AA\ filter.
The red line is the best fitting stellar template smeared at
the best fitting velocity dispersion. Hatched regions represent those 
masked out during the fit. The top scale shows the rest frame wavelength
and the derived quantities.
\label{fig:t15i7196}}
\end{inlinefigure}

\begin{inlinefigure}
\begin{center}
\resizebox{\textwidth}{!}{\includegraphics{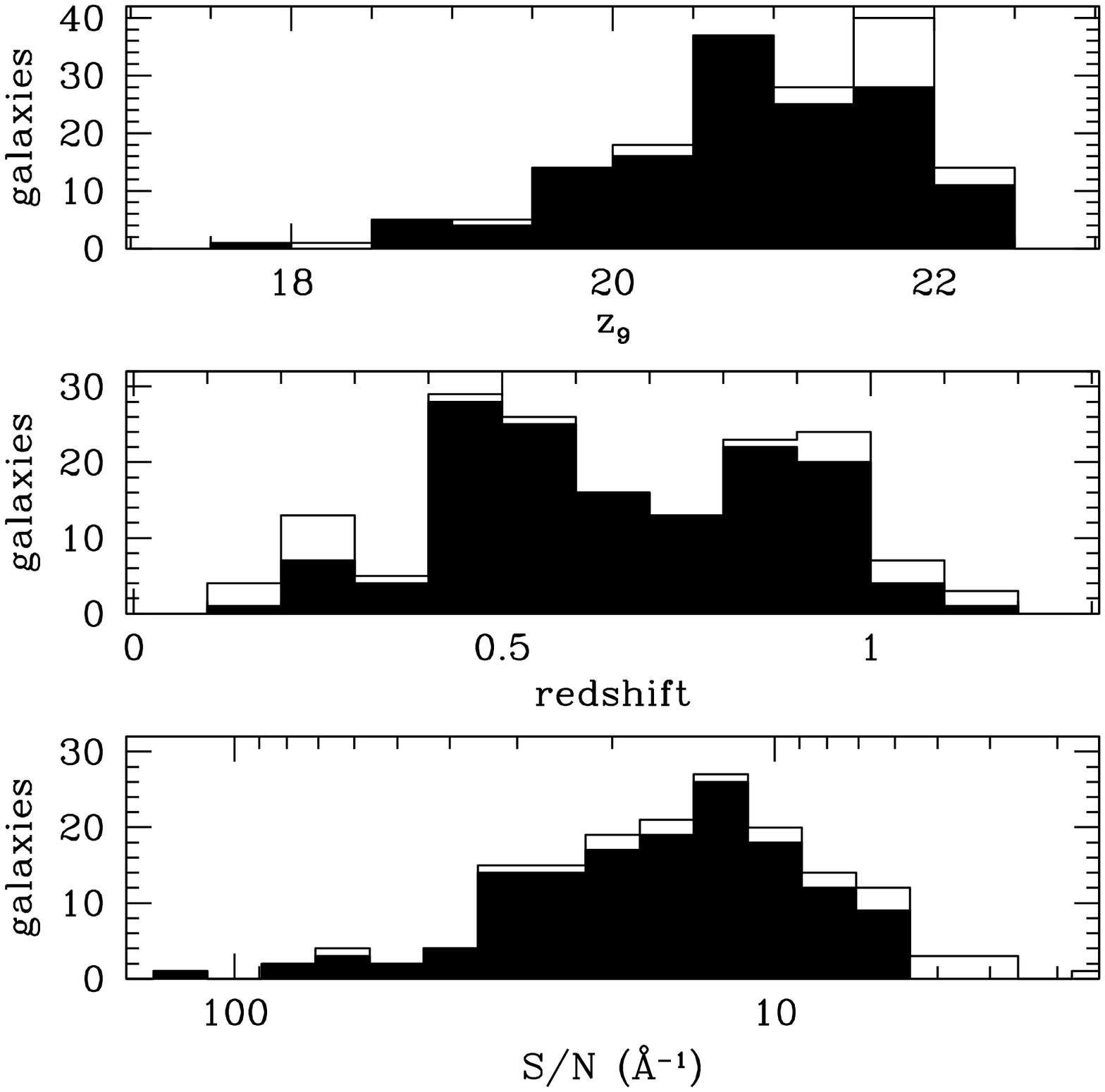}}
\end{center}
\figcaption{Success rate of velocity dispersion measurements. The
upper panel shows the redshift distribution of early-type galaxies in
the spectroscopic sample (empty histogram) and those with measured
velocity dispersions (solid histogram). The middle panel shows the
equivalent $z_{\rm 9}$ luminosity distribution of the two samples. The
lower panel shows the $<$S/N$>$ (\AA$^{-1}$;observer frame)
distribution of the two samples (note the inverted logarithmic
x-axis).
\label{fig:scomplete}}
\end{inlinefigure}

\begin{inlinefigure}
\begin{center}
\resizebox{\textwidth}{!}{\includegraphics{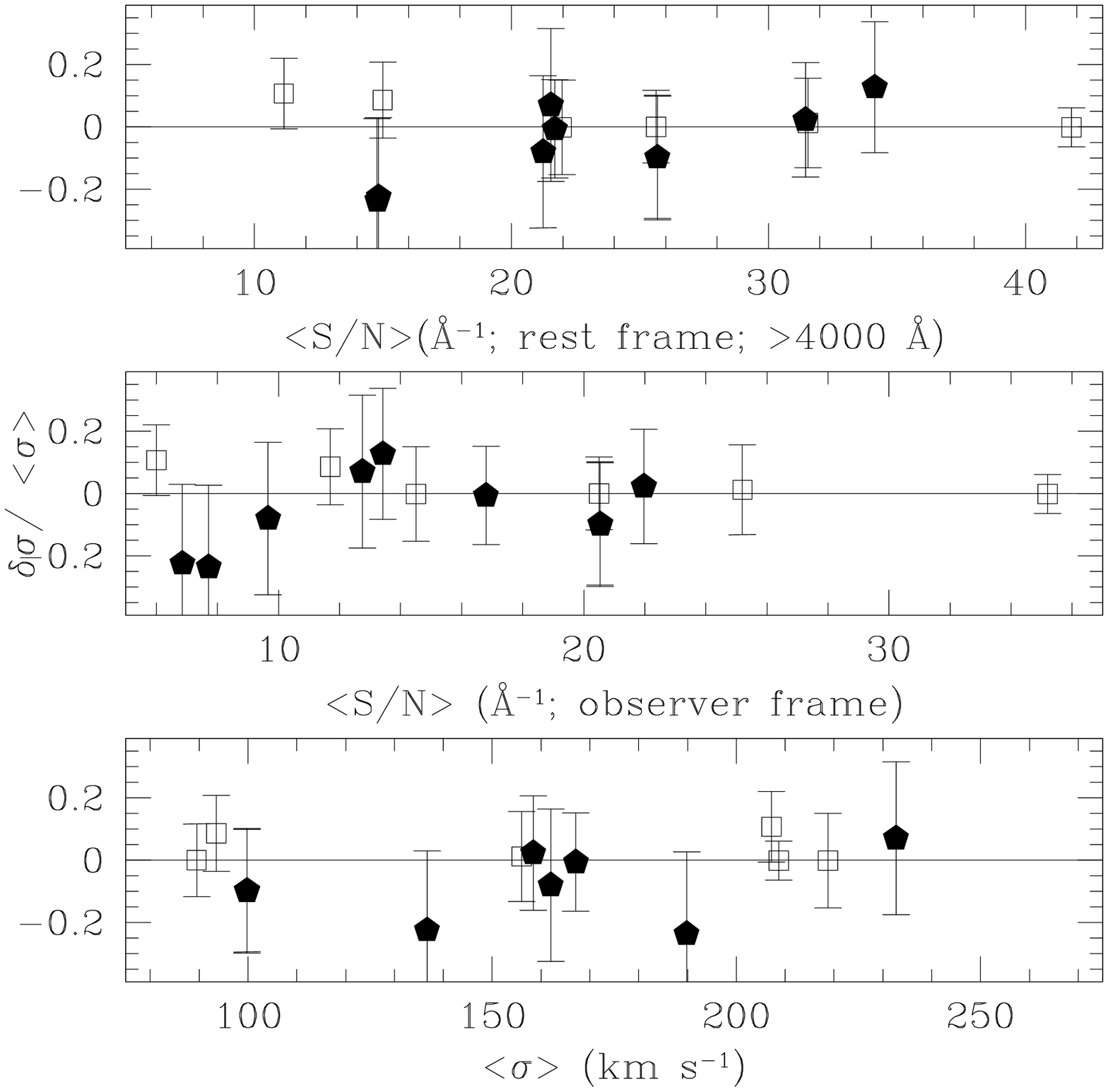}}
\end{center}
\figcaption{Velocity dispersion tests: I. Solid points represent
comparisons with LRIS measurements from van Dokkum \& Ellis (2003), open
points represent internal DEIMOS comparisons for objects observed on
different mask and slitlets configurations.
\label{fig:acfrsigma}}
\end{inlinefigure}

\begin{inlinefigure}
\begin{center}
\resizebox{\textwidth}{!}{\includegraphics{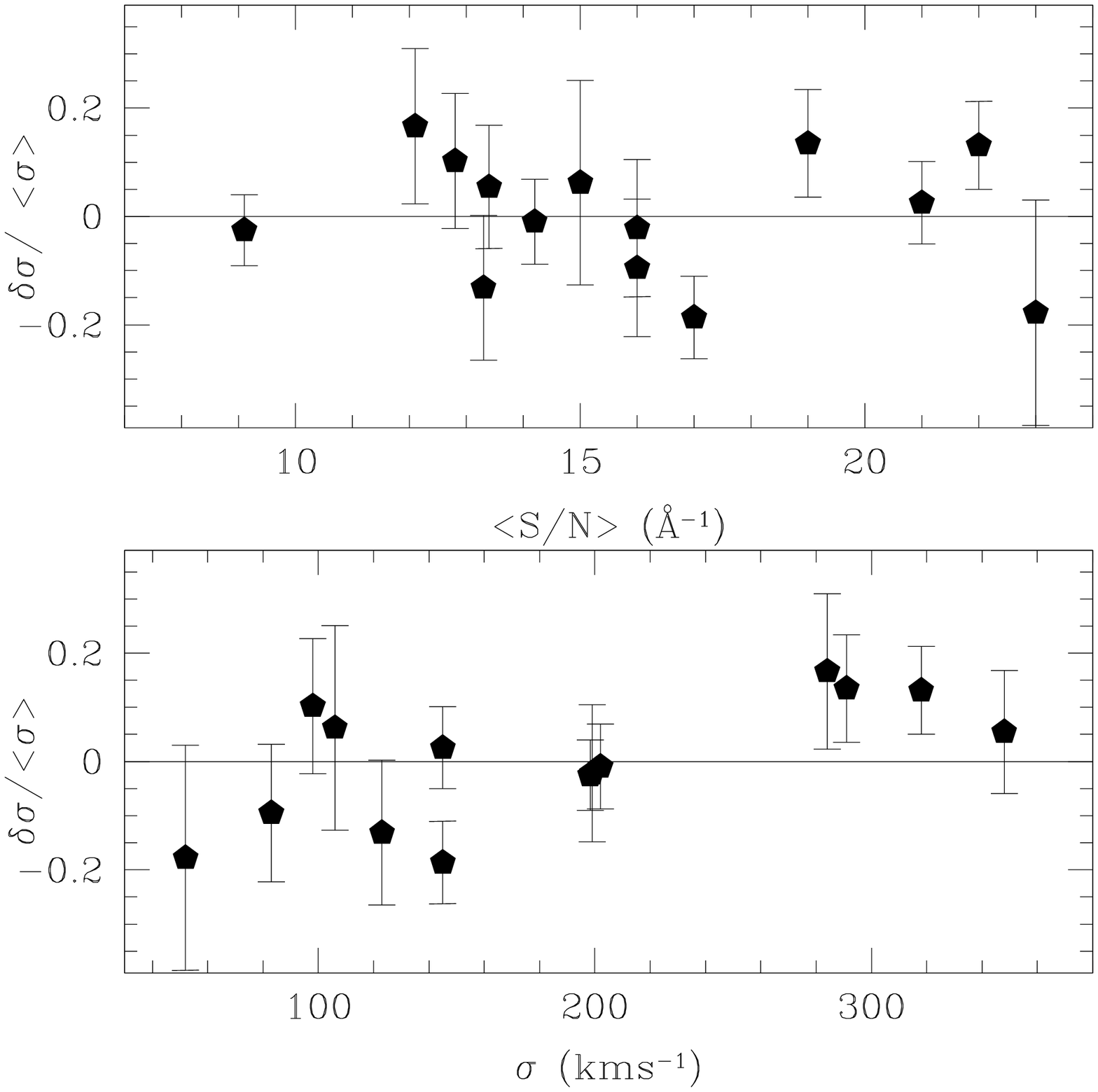}}
\end{center}
\figcaption{Tests on velocity dispersions II. Comparison between
velocity dispersions obtained with two idependent fitting methods
(Treu et al. 2001a and van Dokkum \& Franx 1996).
\label{fig:acfrsigma2}}
\end{inlinefigure}

\begin{inlinefigure}
\begin{center}
\resizebox{\textwidth}{!}{\includegraphics{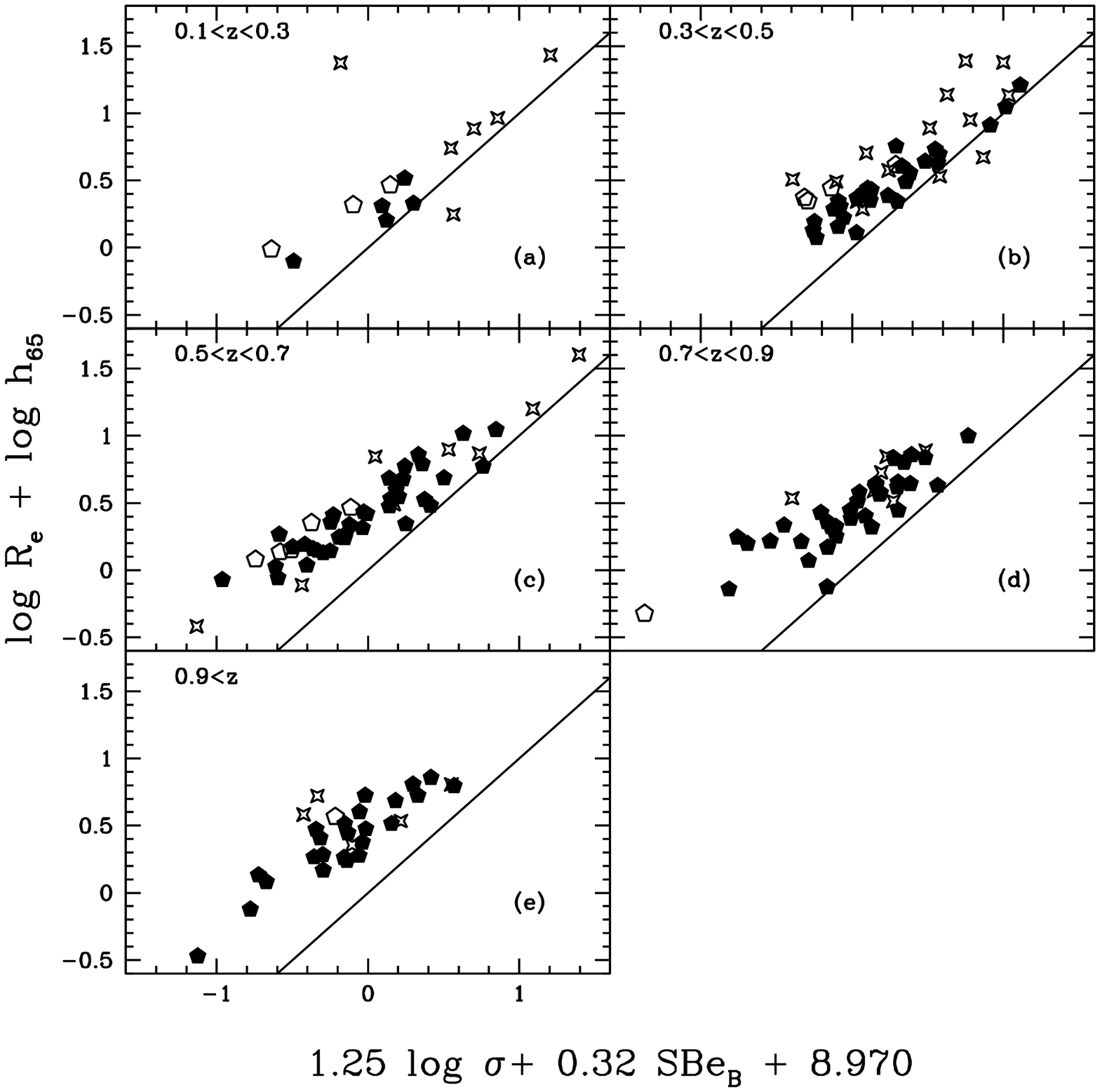}}
\end{center}
\figcaption{Location of survey galaxies in FP-space (the projection
corresponds approximately to an-edge on view of the local FP) binned
in redshift. Solid pentagons represent E+S0 galaxies with
$\sigma>=100$ \kms, open pentagons represent E+S0 galaxies with
$\sigma<100$ \kms, while starred squares represent Sa+b galaxies. The
local relationship is shown as a solid line.
\label{fig:FPev5Bn}}
\end{inlinefigure}

\begin{inlinefigure}
\begin{center}
\resizebox{\textwidth}{!}{\includegraphics{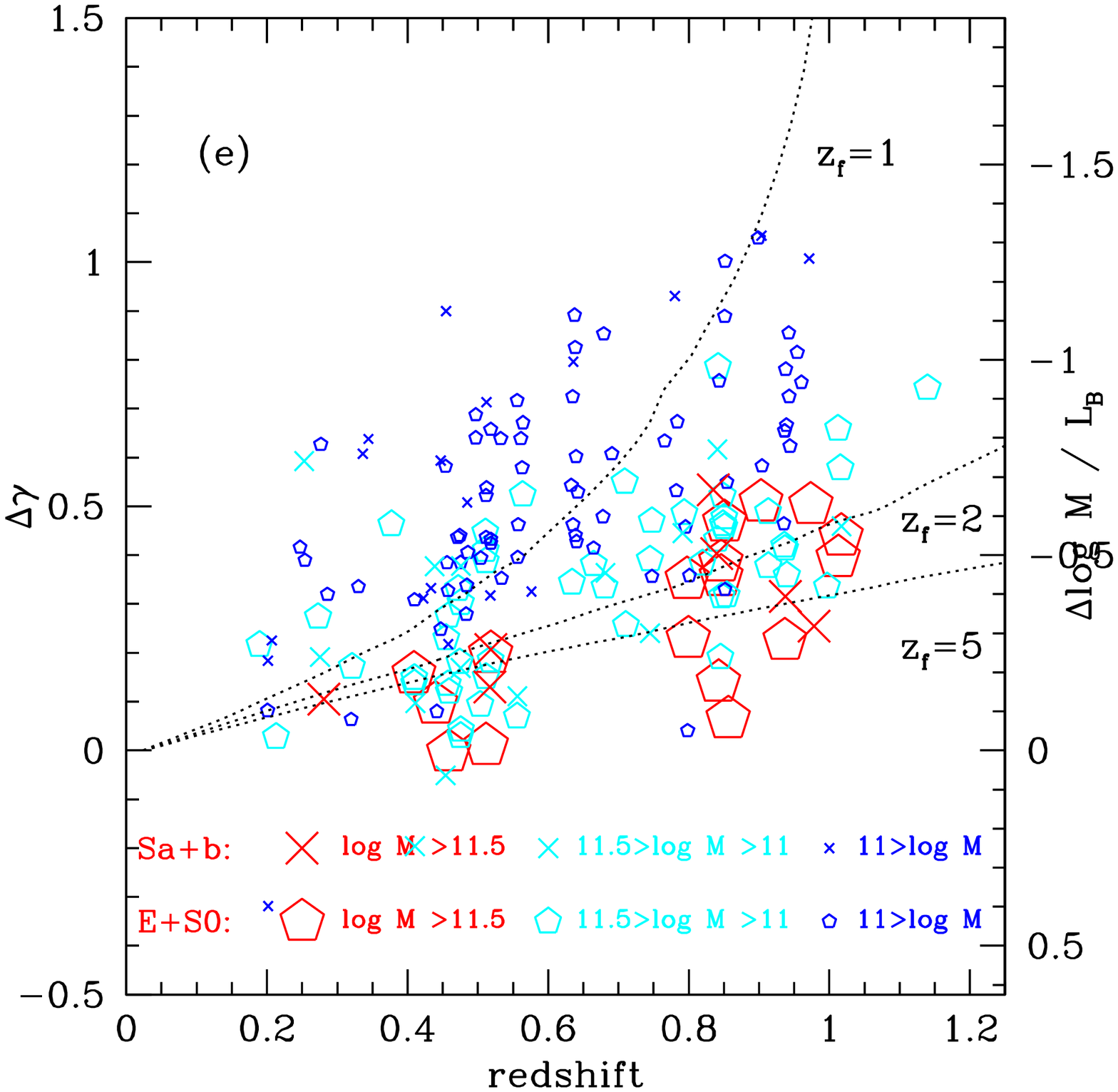}}
\end{center}
\figcaption{Offset from the local FP for the early-type and
early-spiral galaxies, coded by mass (bigger symbols represent bigger
dynamical masses). Notice that the offset from the local relationship
and the scatter decrease with mass.
\label{fig:FPBpJ}}
\end{inlinefigure}

\begin{inlinefigure}
\begin{center}
\resizebox{\textwidth}{!}{\includegraphics{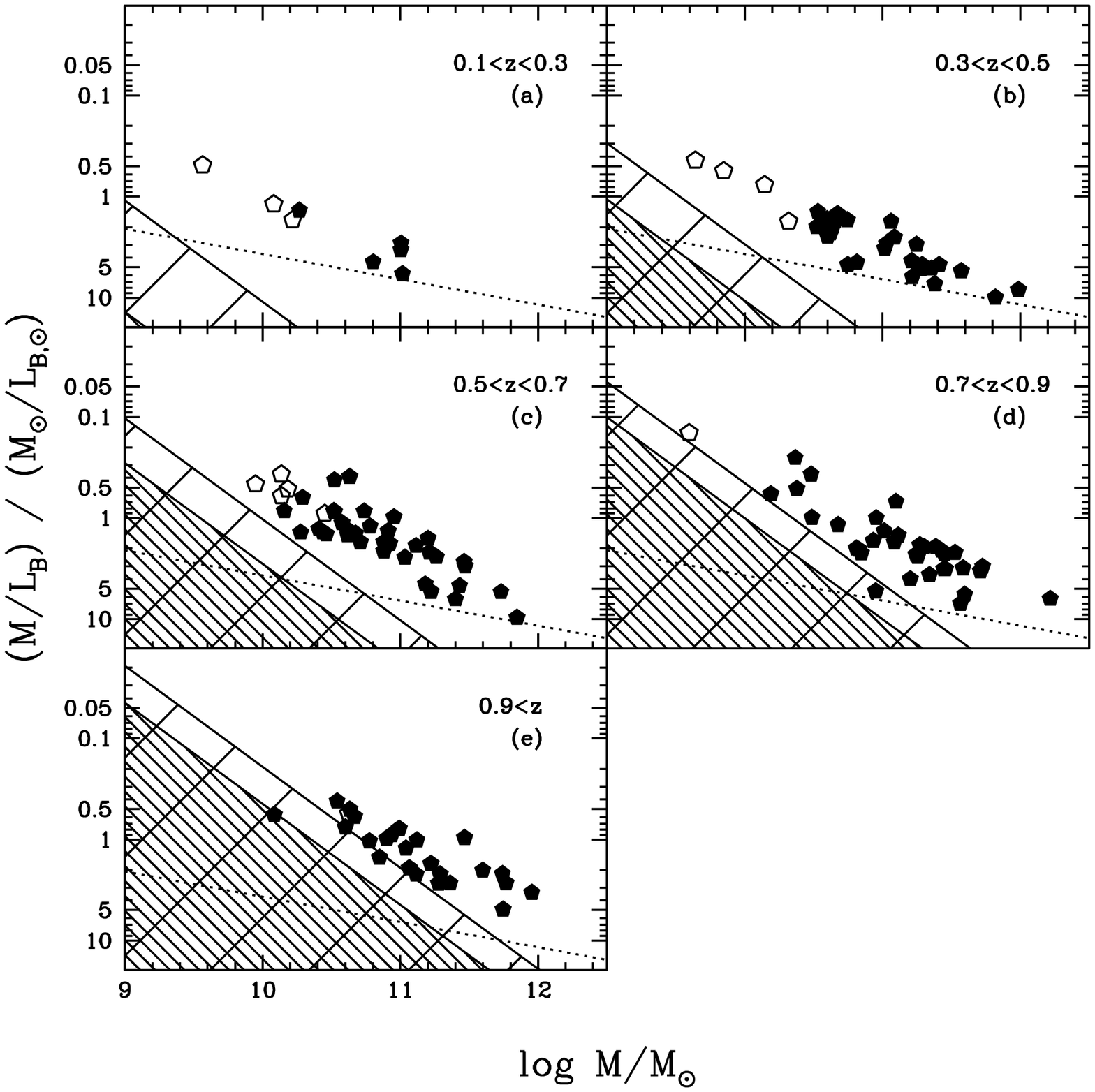}}
\end{center}
\figcaption{Effects of magnitude selection on the determination of the
FP. The panels and symbols are as in Figure~\ref{fig:FPev5Bn} but the
FP is now projected on the M-M/L plane. The local relation is shown as
a dotted line for comparison. Hatched regions are excluded by our
$z_{\rm 9}<22.43$ magnitude limit. The more densely hatched regions
indicate the selection limit applicable for the lower end of the
redshift bin, the more lightly hatched one to the upper end (assumed
at $z$=1.1 for panel e). The plots are for illustrative purposes only;
the limits are approximate to a few hundreds of a magnitude because a
fixed $k$-color correction was adopted and because the model
photometry used to construct M/L$_{\rm B}$ is somewhat different from
the SExtractor photometry in the selection process. The true selection
criteria in the $z_{\rm 9}$ band are faithfully implemented in our
Montecarlo-Bayesan method.
\label{fig:FPev5Bk}}
\end{inlinefigure}

\begin{inlinefigure}
\begin{center}
\resizebox{\textwidth}{!}{\includegraphics{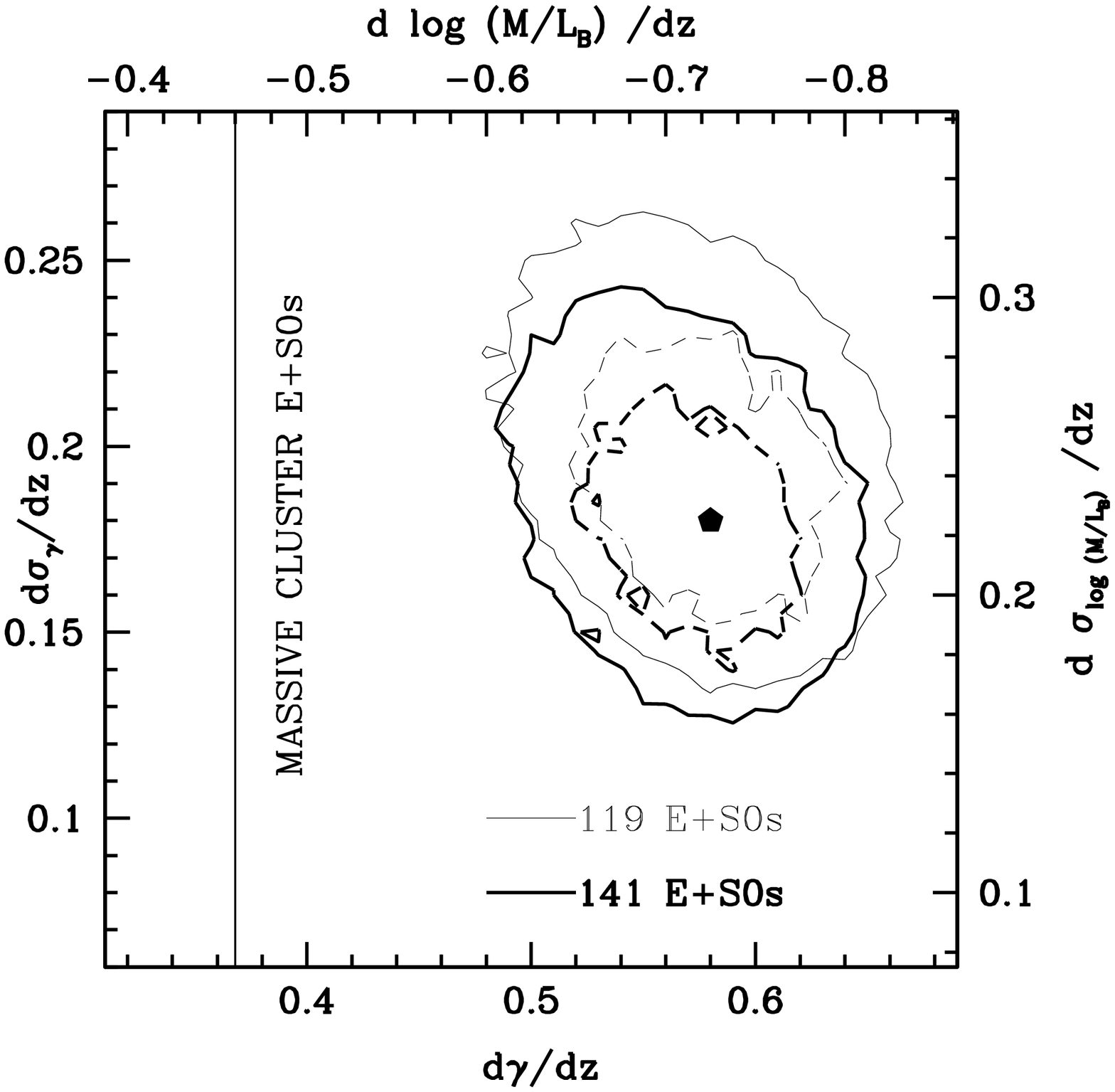}}
\end{center}
\figcaption{
\label{fig:Monte1} Probability contours for the evolution of the 
intercept $d\gamma/dz$ and maximal evolution of the scatter
$d\sigma_{\gamma}/dz$ (dashed=68\%; solid=95\%). The maximum of the
probability distribution is indicated by the solid pentagon. Thin
contours represent the same probability levels for a smaller sample of
119 E+S0s, where 21 S0s have been excluded as possible early-spirals
contaminants, see Section~\ref{ssec:morph}. The corresponding values
in terms of $M/L$ are shown on the top and right axis.}
\end{inlinefigure}

\begin{inlinefigure}
\begin{center}
\resizebox{\textwidth}{!}{\includegraphics{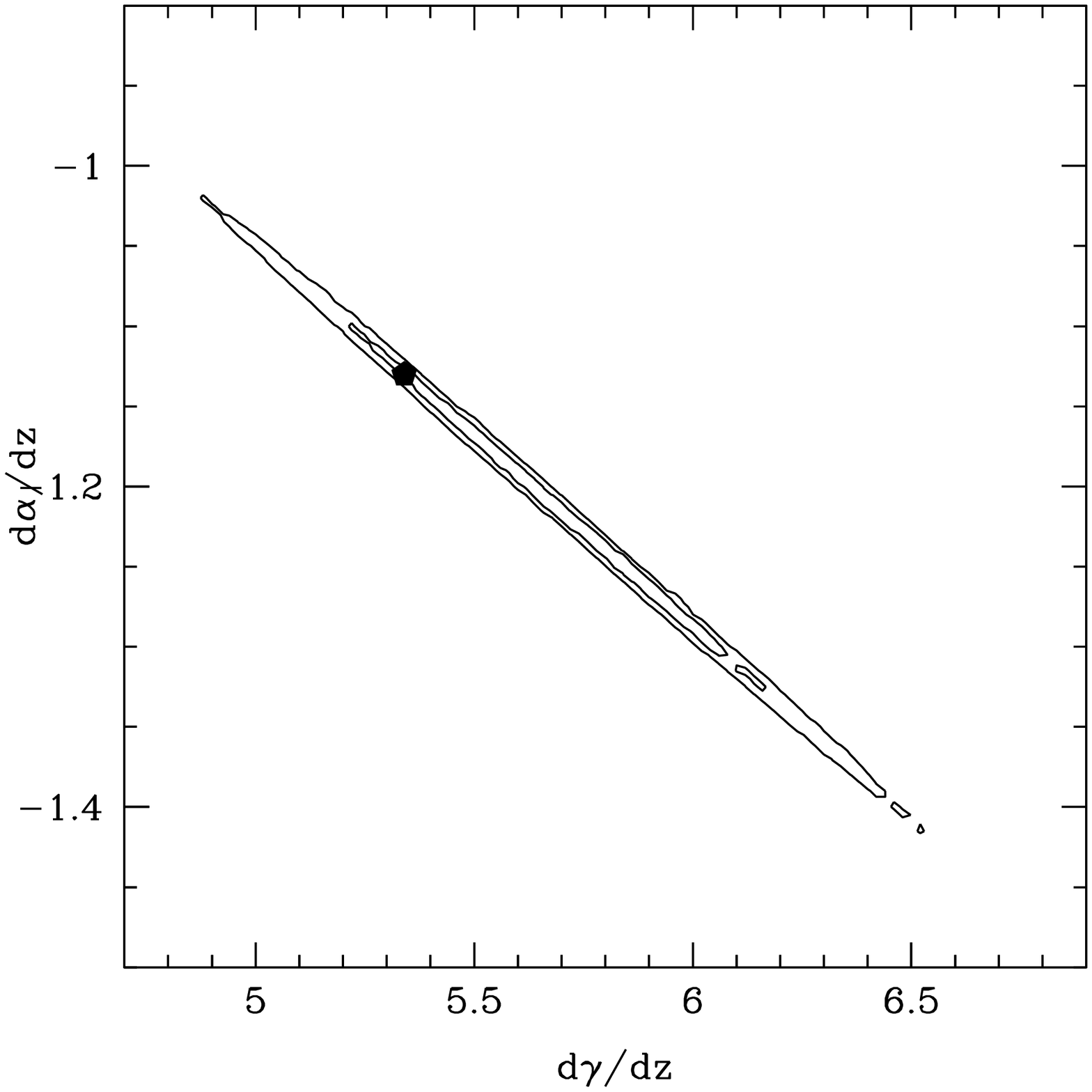}}
\end{center}
\figcaption{
\label{fig:Monte2} Mass-dependent evolution of the mass-to-light ratio:
joint probability contours (68\% and 95\%) for evolution of the slope 
$\alpha$ and the intercept $\gamma$ of the FP, marginalized over the 
evolution of the scatter, and assuming $\alpha(z)+10\beta(z)+2=0$.}
\end{inlinefigure}

\begin{inlinefigure}
\begin{center}
\resizebox{\textwidth}{!}{\includegraphics{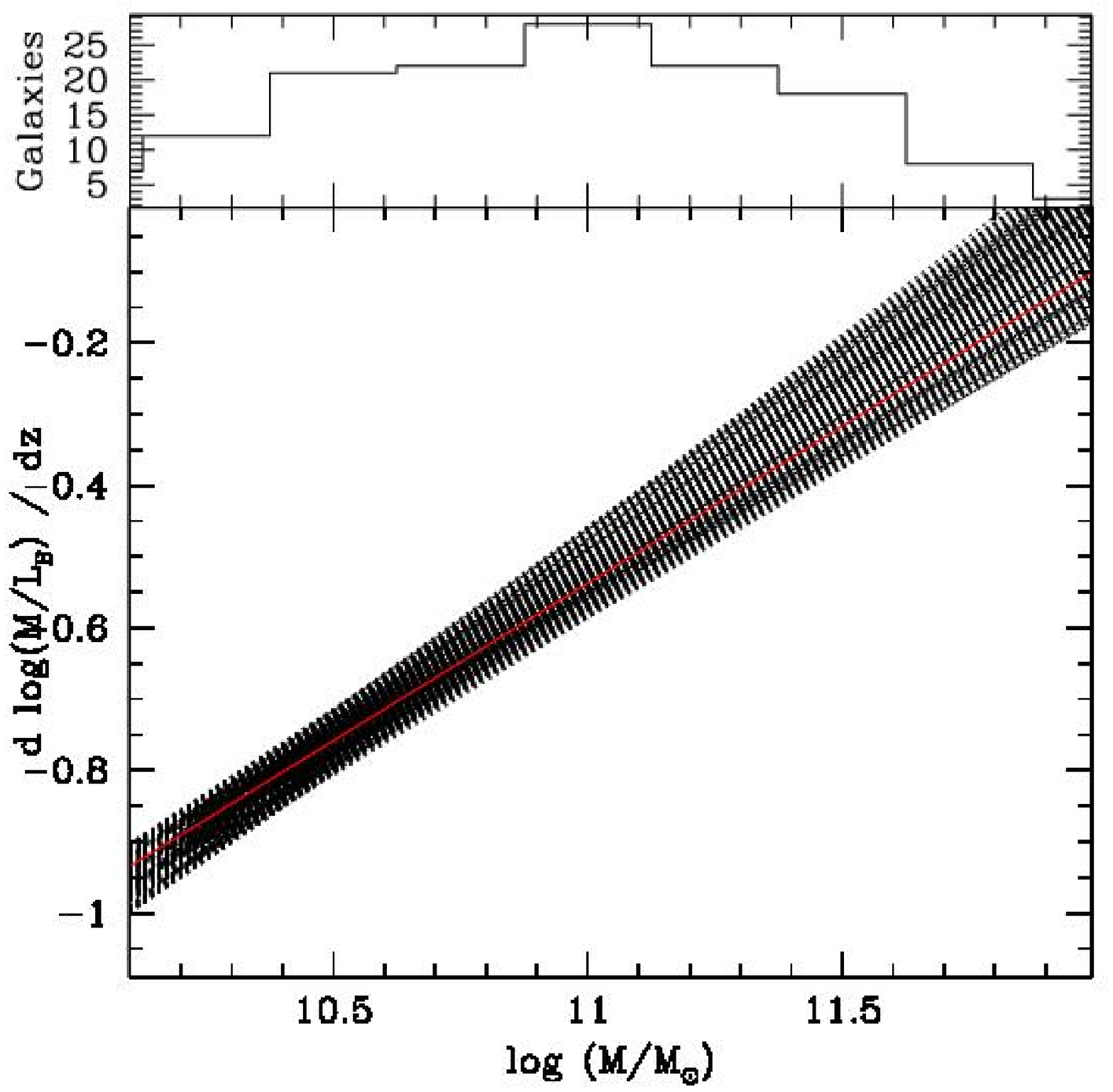}}
\end{center}
\figcaption{
\label{fig:Monte3}Bottom panel: the shaded area represents the 
68\% probability contour for the evolution of $M/L$ as a function of
$M$, obtained from Figure~\ref{fig:Monte2} with
Equation~\ref{eq:massev}. The solid red line corresponds to the best
fitting model. Top panel: distribution of dynamical masses for our
sample.}
\end{inlinefigure}

\begin{inlinefigure}
\begin{center}
\resizebox{\textwidth}{!}{\includegraphics{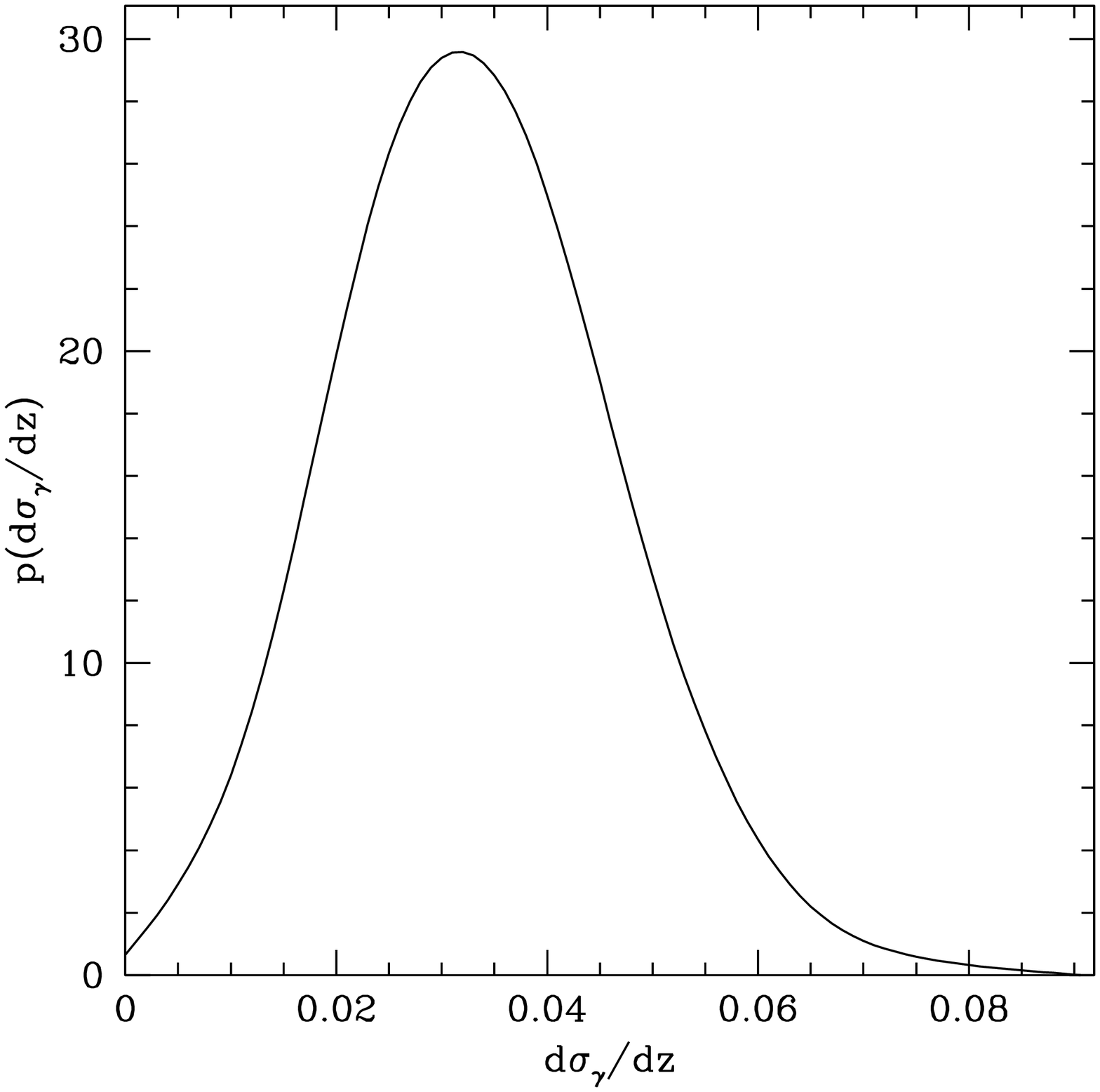}}
\end{center}
\figcaption{
\label{fig:Monte4}Probability distribution for the evolution of the scatter 
(in $\log R_{\rm e}$) of the FP, marginalized over changes in slopes
and intercept. The scatter evolves as
$d\sigma_\gamma/dz=0.032\pm0.012$ (68\% limits).}
\end{inlinefigure}

\begin{inlinefigure}
\begin{center}
\resizebox{\textwidth}{!}{\includegraphics{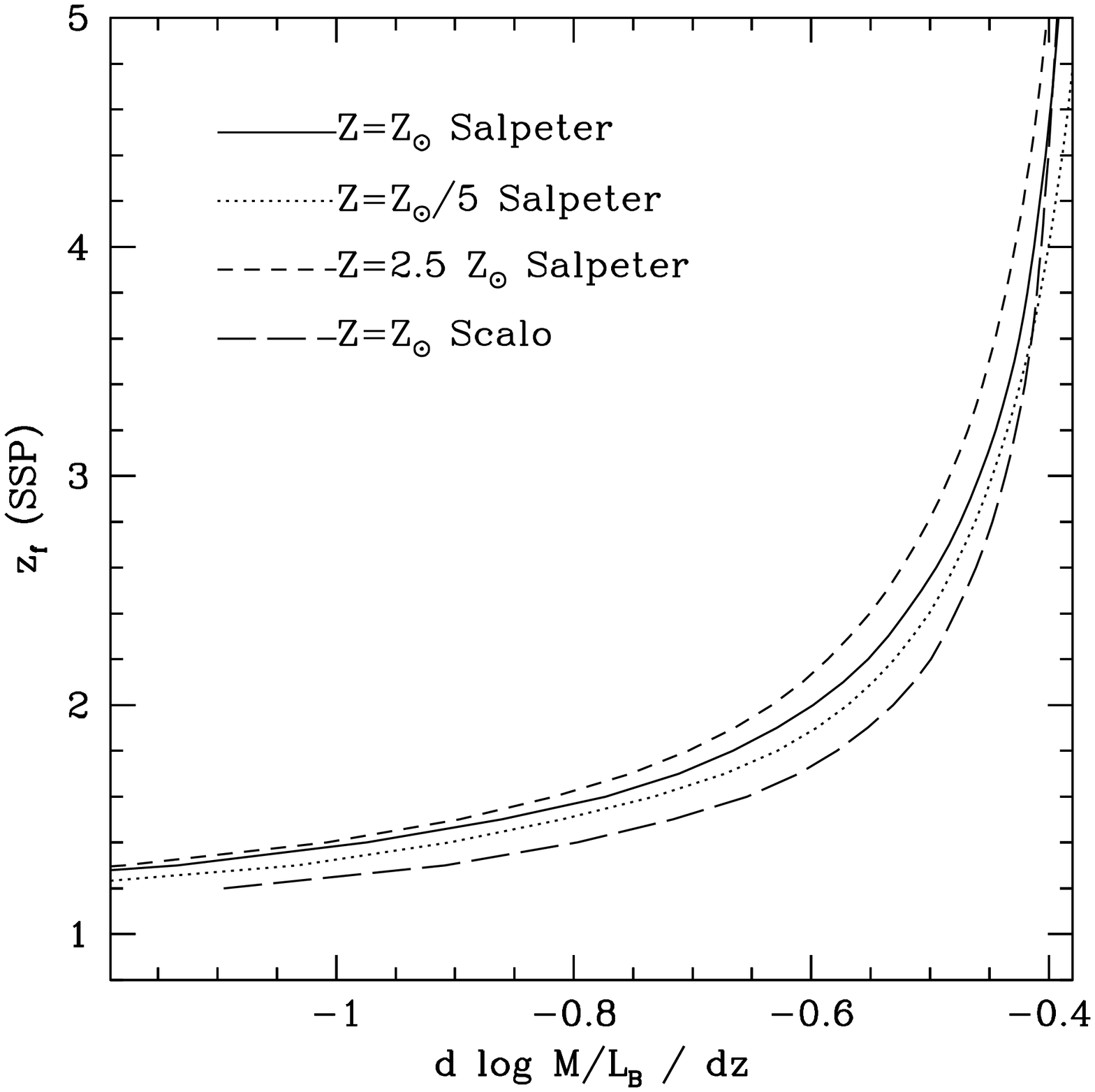}}
\end{center}
\figcaption{
\label{fig:zf} Redshift of formation as a function of evolution of the 
effective mass to light ratio for a set of simple stellar population
models.}
\end{inlinefigure}

\begin{figure*}[t]
\begin{center}
\resizebox{\textwidth}{!}{\includegraphics{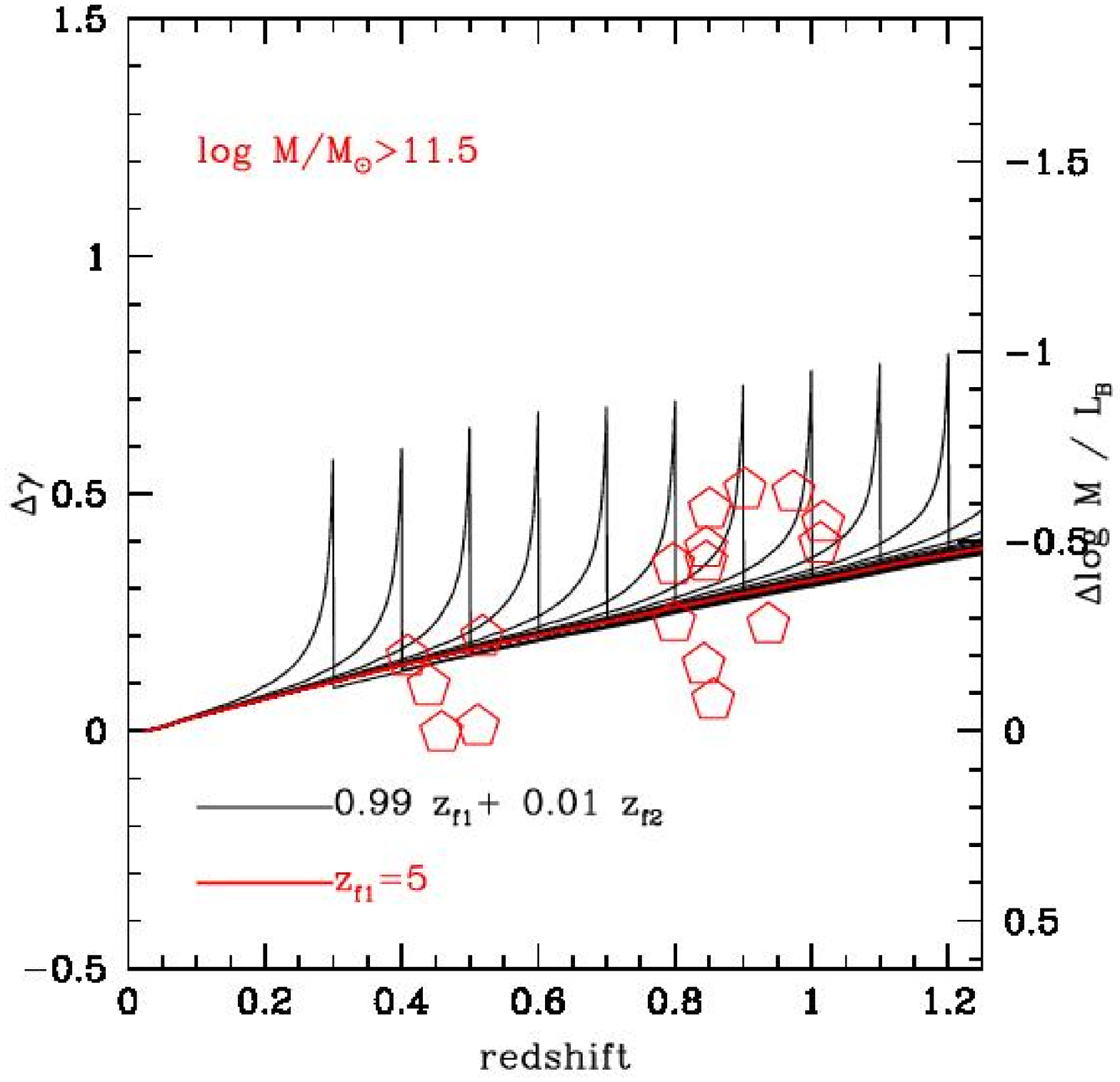}\includegraphics{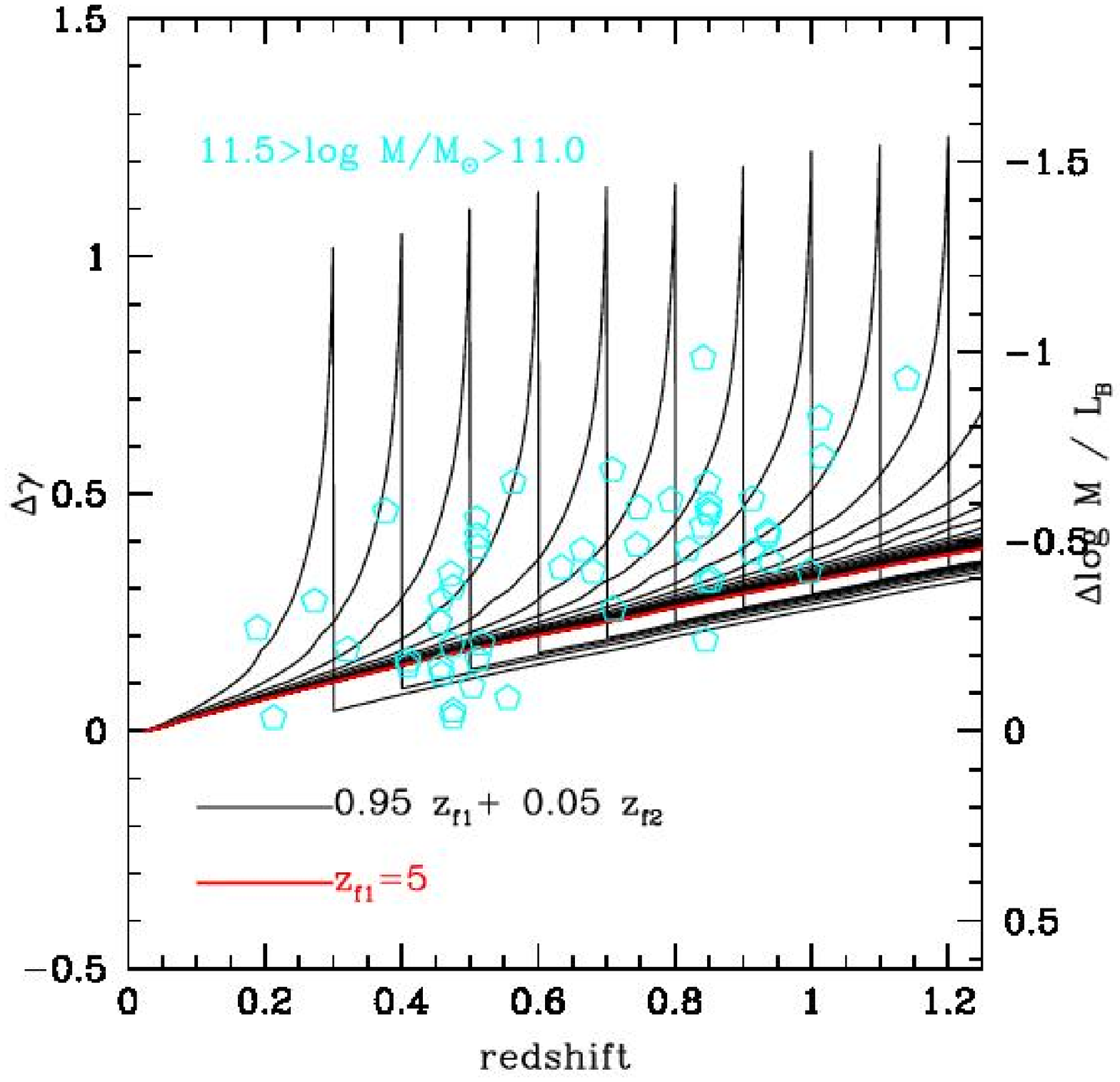} \includegraphics{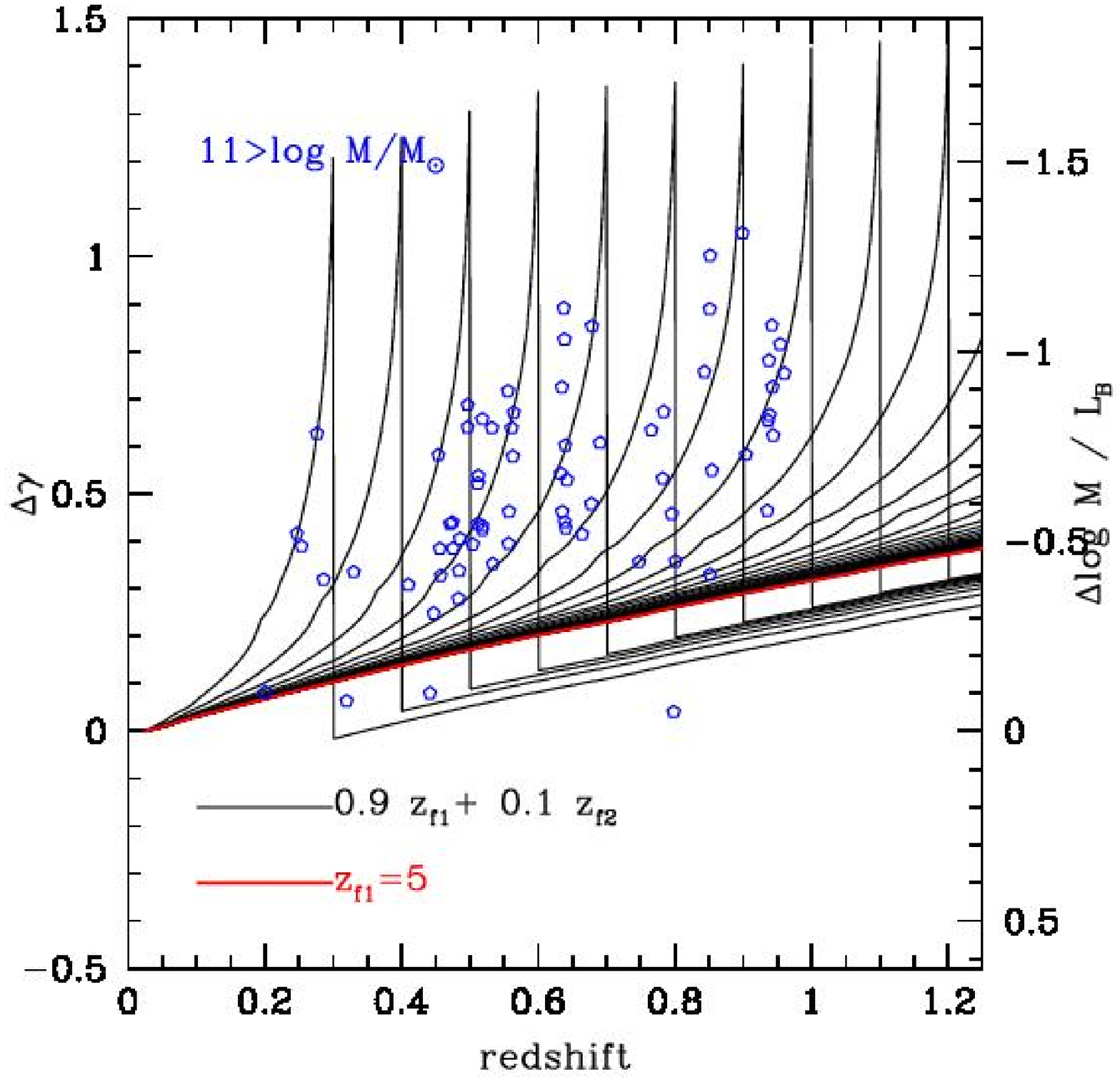}}
\end{center}
\figcaption{
\label{fig:zf2} The solid black lines show the evolution of the
 mass-to-light ratio for a family of two-bursts models, where the
majority of the stellar mass is formed at $z_{f1}=5$ and the rest
(1-5-10\% as indicated on the plots) is formed at $z_{f2}<z_{f1}$
(secondary bursts are shown in steps of $dz=0.1$, in the range
$z=0.3-3$). For illustration the evolution of the mass-to-light ratio
of the old component is also shown as a thick solid red line. The
observed offset for our sample of E+S0 galaxies (coded by mass as in
previous figures), is also shown for comparison. Note that this simple
model covers the observed range of parameters, demonstrating that a
simple two burst model is sufficient to reproduce the observations.}
\end{figure*}

\begin{inlinefigure}
\begin{center}
\resizebox{\textwidth}{!}{\includegraphics{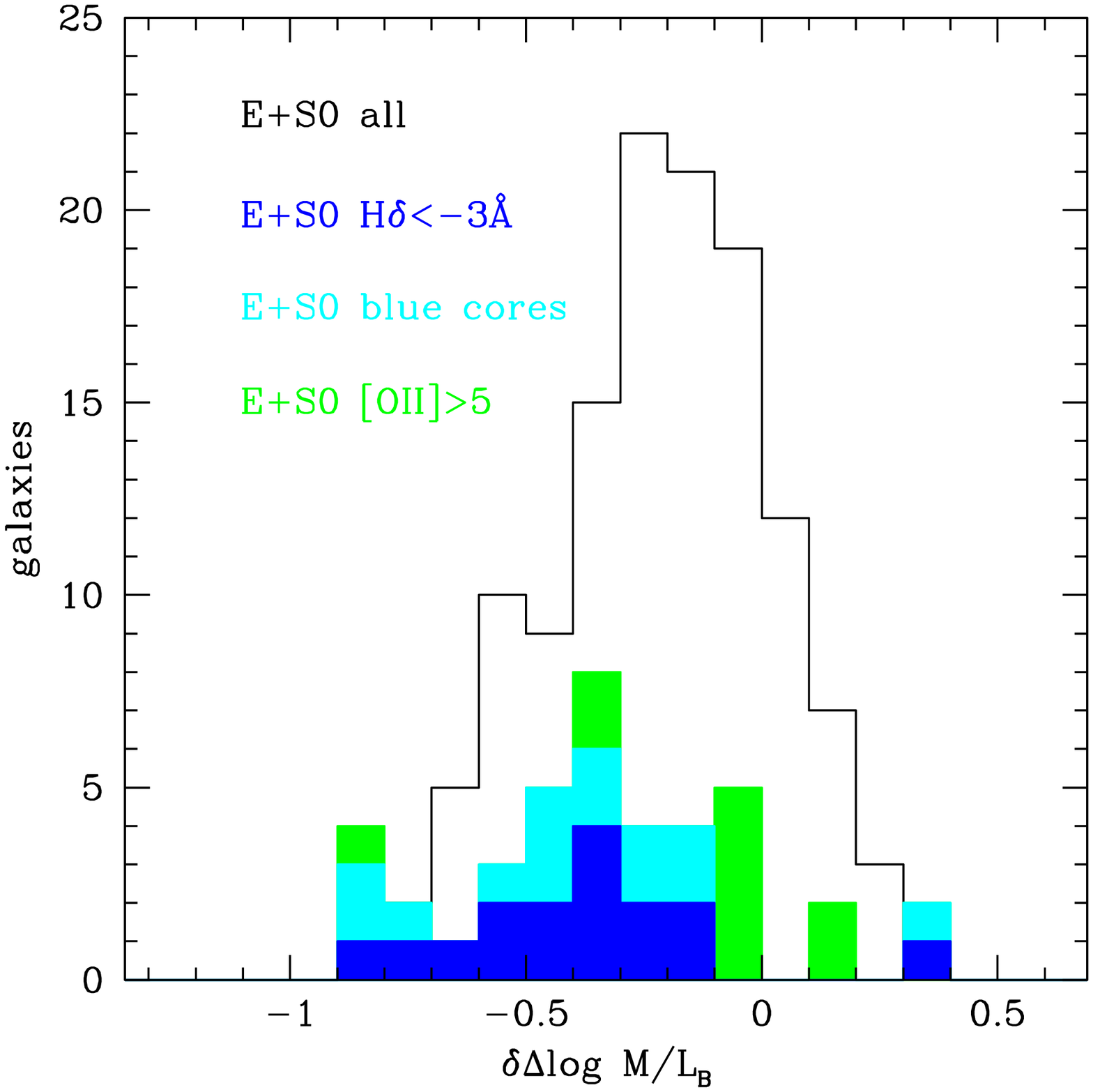}}
\end{center}
\figcaption{
\label{fig:DelBC} {Distribution of offsets of the mass-to-light ratio 
evolution relative to that defined by van Dokkum \& Stanford for
massive cluster galaxies for all E+S0s, those with strong H$\delta$
blue cores and [\ion{O}{2}] emission.}}
\end{inlinefigure}

\begin{inlinefigure}
\begin{center}
\resizebox{\textwidth}{!}{\includegraphics{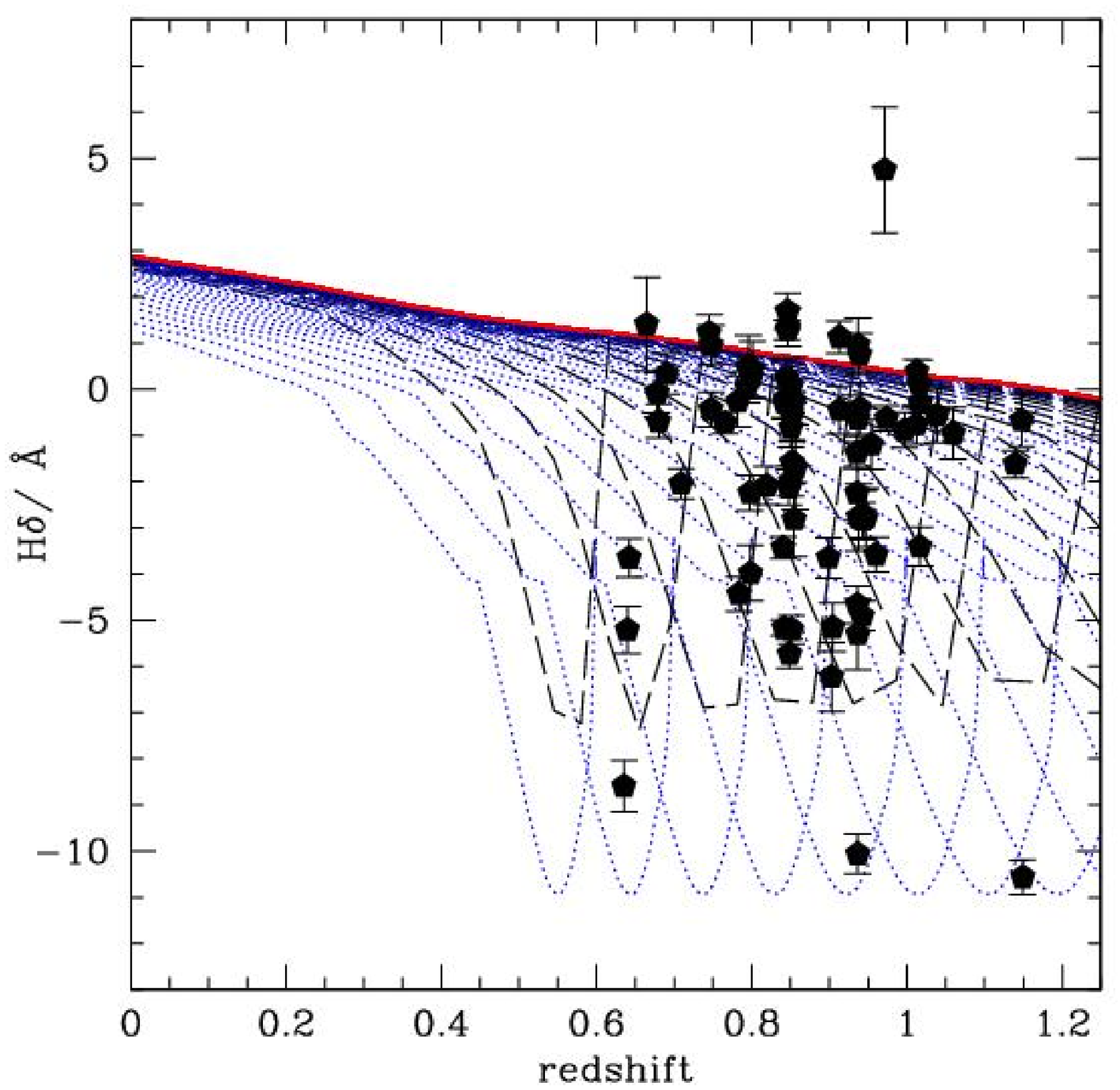}}
\end{center}
\figcaption{H$\delta$ as a function of redshift for a family of two
bursts models, consisting of a main burst forming 90\% of the stellar
mass at $z_{f1}=3$, plus a secondary burst forming 10\% of the stellar
mass at $z_{f2}<z_{f1}$ (lines as in Figure~\ref{fig:zf2}; models from
Bruzual \& Charlot 2003; solar metallicity). Measurements of H$\delta$
for our sample of E+S0s galaxies are shown as filled pentagons.
\label{fig:Hd2}}
\end{inlinefigure}

\begin{inlinefigure}
\begin{center}
\resizebox{\textwidth}{!}{\includegraphics{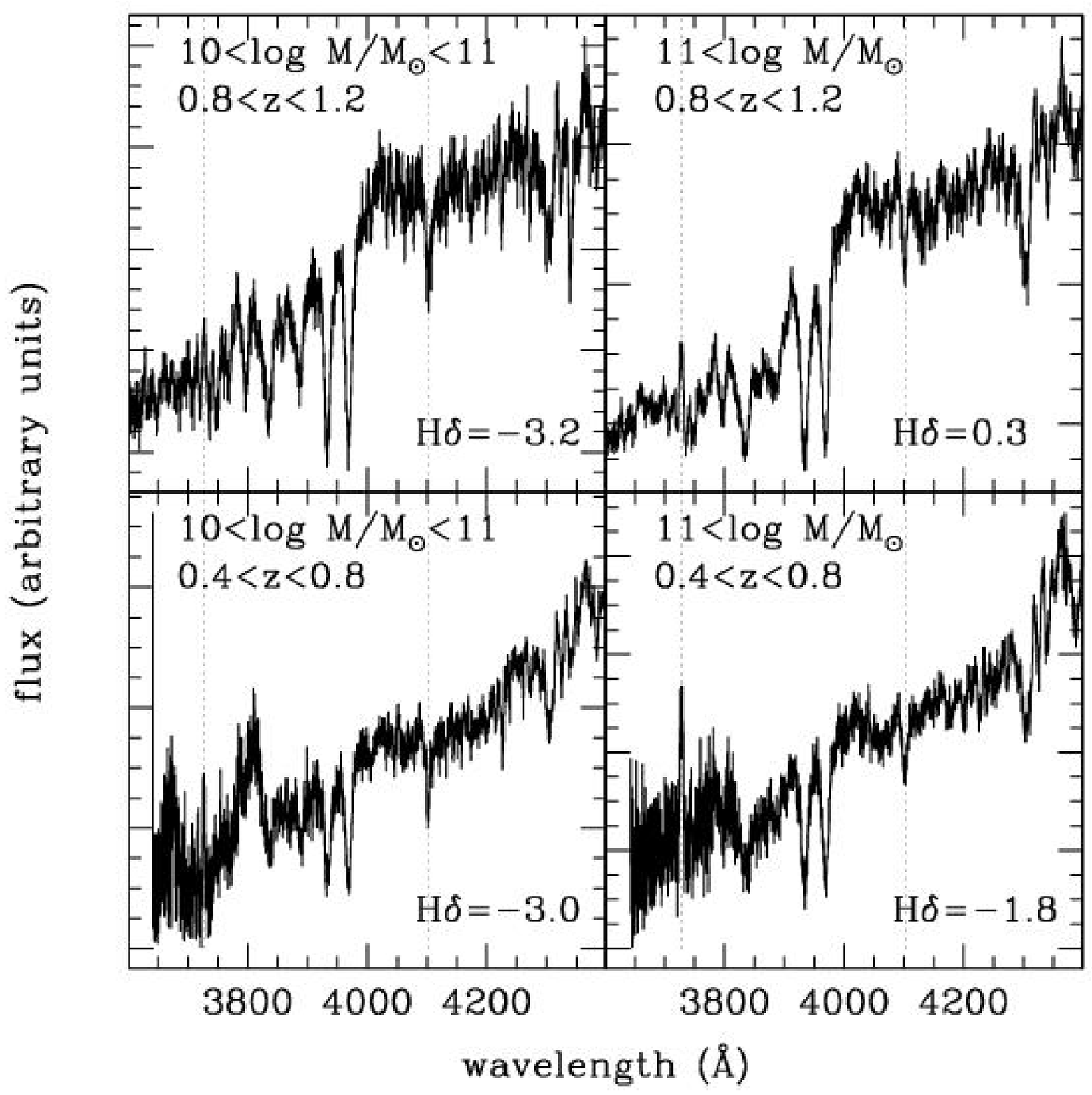}}
\end{center}
\figcaption{Co-added spectra of E+S0 galaxies, binned in mass and
redshift. H$\delta$ in absorption is seen to be stronger for smaller
mass systems, consistent with younger stellar ages than their massive
counterparts.
\label{fig:coaddev2}}
\end{inlinefigure}

\end{document}